\documentclass[twocolumn,amsmath,amssymb,
aps, showpacs,graphicx]{revtex4-2}

\usepackage{graphicx}
\usepackage{dcolumn}
\usepackage{bm}
\usepackage{lineno, hyperref}
\usepackage{natbib}
\begin{document}

\title{Antiferromagnetic ordering and excitonic pairing in the AA-stacked bilayer graphene}
\author{V. Apinyan} 
\altaffiliation[e-mail:]{v.apinyan@intibs.pl}
\author{T. K. Kope\'{c}} 
\affiliation{Institute of Low Temperature and Structure Research, Polish Academy of Sciences\\ PO. Box 1410, 50-950 Wroc\l{}aw 2, Poland 
}%

\begin{abstract}
	
In the present paper, we describe the antiferromagnetic and excitonic correlations in the AA-stacked bilayer graphene. We consider the applied external electric field potential to the structure which leads to the electronic charge imbalance between the layers in the system. By using the generalized two-layer Hubbard Hamiltonian, we consider different particle filling regimes in the layers. We calculate the important energy scales in the system and we establish the conditions for the appearance of the antiferromagnetic order in the system. We consider both large and small Coulomb interaction limits in the layers and the effect of the electric field potential on the calculated order parameters. We discuss the coexistence of antiferromagnetism and excitonic phases and we show that they can coexist only in the regime away from half-filling. In the case away from the half-filling, we show the existence of a critical value $U_{\rm C}$ of the Coulomb interaction potential at which we establish the transition from the single-valued to the triple valued excitonic states, governed by the strong electronic reconfiguration, in the system. The zero-temperature limit is considered in the problem.
	
\end{abstract}



\maketitle

\section{Introduction}
%
The electronic properties of the AB-stacked and AA-stacked bilayer graphene (BLG) structures, their optical characteristics and excitonic phases have been the subject of many experimental and theoretical works \cite{cite_1, cite_2, cite_3, cite_4, cite_5, cite_6, cite_7, cite_8, cite_9, cite_10, cite_11, cite_12, cite_13, cite_14, cite_15, cite_16, cite_17, cite_18, cite_19, cite_20, cite_21, cite_22, cite_23, cite_24, cite_25, cite_26, cite_27, cite_28, cite_29, cite_30, cite_31}. Beside the fact that the AB-BLG is very promising for technological applications, the AA-BLG shows intriguing properties when exposed to the external electric field \cite{cite_1, cite_2, cite_5, cite_6, cite_7, cite_16, cite_27, cite_28, cite_30, cite_31} and a series of researches was done to fabricate such structures \cite{cite_32, cite_33, cite_34, cite_35, cite_36}.   

The multilayer graphene systems appear to be ideal candidates for studying the exciton formation and condensation phenomena. The excitonic pairing in bilayer graphene systems, has been reported in many studies \cite{cite_19, cite_20, cite_21, cite_22, cite_23, cite_24, cite_25, cite_26, cite_27}. In turn, this can provide the new possibilities to construct a new category of optoelectronics, based on the excitonic qubit-functionality \cite{cite_37, cite_38}. The manipulation of the excitonic particles, instead of the individual photons (as it was done till now), increase the energy consumption efficiency and leads to the nano-environments with the smallest possible energy-dissipation \cite{cite_39, cite_40}. Although the experimental observation of the excitonic insulator state in BLG remains difficult (because the electron-hole recombinations destroy the excitonic binding states), the spin-triplet excitonic states can be detected experimentally \cite{cite_41} by the spin-transport measurements, \cite{cite_42} and are important for spintronics applications \cite{cite_43}. However, for detecting the spin-superfluidity one needs the spin-polarization in the system.   
 
A series of works has been consecrated to study the coexistence of spin-polarized states with the excitonic states in the AA-BLG system \cite{cite_27, cite_28, cite_29, cite_44, cite_45, cite_46, cite_47, cite_48, cite_49}.
The existence of the antiferromagnetism is proved to be important for the spin neutral edge states which leads to the negligible stray fields in the system and are robust against the magnetic field perturbation \cite{cite_8, cite_15}.  

The strength of the Coulomb interaction in graphene and graphite is accurately determined in Ref.\cite{cite_46} by first-principle calculations. A critical value of the Coulomb interaction was found \cite{cite_47, cite_48, cite_49, cite_50}, at which a transition occurs from nonmagnetic semi-metallic to the antiferromagnetic insulator states in graphene by using quantum Monte Carlo and finite-size scaling for the Hubbard model. The Dependence of the antiferromagnetic condensate state on the on-site electron-electron interaction is examined in Ref.\cite{cite_45}. Moreover, the coexistence of the antiferromagnetic and excitonic states in the AA-BLG system, away from the half-filling regime, has not been examined yet. 

In the present paper, we calculate the antiferromagnetic and excitonic order parameters in the AA-BLG system (see, in Fig.~\ref{fig:Fig_1}), in which the atoms in the upper layer are placed at top of the atoms in the lower layer. Different interaction limits and filling coefficients are considered in the problem. In our case, we consider the inverse filling-factor $\kappa$, thus, the coefficient $\kappa^{-1}$ gives the usual filling coefficient, known in the literature (see, for example, in Refs.\cite{cite_2, cite_27, cite_28, cite_29, cite_44, cite_45}): $\bar{n}_{\eta_{i}}+\bar{n}_{\eta_{j}}=1/\kappa$, where $\bar{n}_{\eta_{i}}$ are the average fermionic densities for the $\eta$-type sublattices ($\eta=A$ or $\eta=B$) and the indexes $i$ and $j$ (denoting different layers in the system) are such that $i\neq j$. We show here that the transition into the antiferromagnetic phase occurs at some critical values $U_{C}$ and $V_{C}$ of the Coulomb interaction parameter $U$ and external gate potential $V$. The excitonic transition also takes place at the same values. This result was missing in the previous works Refs.\cite{cite_2, cite_27, cite_28, cite_29, cite_44, cite_45}, due to the consideration of the half-filling regime only.

Moreover, we show that for the region below the critical values $U_{C}$ and $V_{C}$ only the excitonic singlet phase exists, while above the critical values the order parameters $\Delta_{\uparrow}$ and $\Delta_{\downarrow}$ pass into triple valued states. In other words, this is similar to the spin-Zeeman effect \cite{cite_51} or the Stark-Lo Sudo type splitting, mediated by the external electric field applied to the system. The triple valued nature of the antiferromagnetic order parameter, for the case of $\kappa=1$, persists in the whole regions $U>U_{C}$ or $V>V_{C}$, while in the regions $U<U_{C}$ and $V<V_{C}$ the antiferromagnetic order parameter vanishes. 

For the case of half-filling, the antiferromagnetic state vanishes completely, while the single-valued excitonic states exist for all values of the applied electric field potential $V$. We show that the AFM and excitonic states can coexist only in the regime away from the half-filling and the existence of the antiferromagnetism in the AA-BLG is strongly related with the appearance of the critical points $U_{C}$ and $V_{C}$. The calculated energy scale of the AFM order parameter $\Delta_{\rm AFM}$ is much larger than for the excitonic order parameter $\Delta_{\sigma}$, i.e., $|\Delta_{\rm AFM}|\gg|\Delta_{\sigma}|$, which brings an idea about the fundamental energy scales in the system. All important physical parameters in the system such as the chemical potential, the average charge density imbalance and the excitonic order parameters $\Delta_{\sigma}$ change their behaviour from single-valued solution to the triple valued ones when passing through the critical values $U_{C}$. Other major effects, found in the present paper, are related to the charge density variations for the regime away from the half-filling and the population inversions between the layers, when passing through the critical values of the Coulomb interaction parameter $U_{C}$ and the applied potential $V$.

The paper is organized as follows: in the Section \ref{sec:Section_2}, we introduce the extended bilayer Hubbard model. In Section \ref{sec:Section_3}, we present the mean-field (MF) decoupling procedure and we obtain the system of coupled, self-consistent, equations which we solve numerically. In Section \ref{sec:Section_4}, we discuss the obtained numerical results, and the physics, related to the coexistence of the antiferromagnetism and excitons. In the Section \ref{sec:Section_5}, we give a conclusion to our paper and, finally, in the Appendix \ref{sec:Section_6}, we present shortly the analytical calculations of the coefficients entering in the system of equations.
%
\section{\label{sec:Section_2} The AA bilayer graphene Hamiltonian}
\begin{figure}
	\begin{center}
		\includegraphics[scale=0.5]{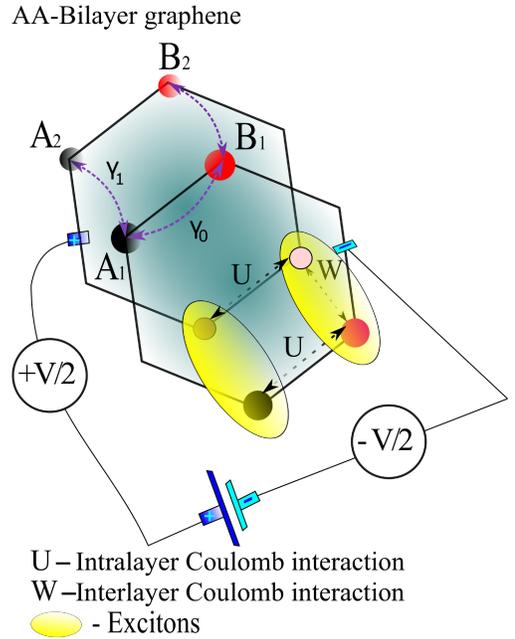}
		\caption{\label{fig:Fig_1}(Color online) The AA type stacked bilayer graphene structure with the applied external electric field potential $V$. The principal physical parameters, entering in the Hamiltonian in Eq.(\ref{Equation_1}), are shown in the picture.}
	\end{center}
\end{figure}

%
The AA type stacked bilayer graphene structure is represented in Fig.~\ref{fig:Fig_1}, where the two layers and the external electric field potential are schematically represented. In this system, the atoms in the top layer lie just above the similar atoms in the bottom layer. As usual, the $\pi_{\rm z}$ electrons participate in the electronic conduction in the layers, while we suppose that the $2p^{1}$-electrons, attached to the carbon atoms, form a $G$-type (shown in the first upper panel, in Fig.~\ref{fig:Fig_2}, above) of the antiferromagnetic lattice (being localized at the positions of the atomic sites, shown in Fig.~\ref{fig:Fig_2}).  

Here, we write the total Hamiltonian of our system of the AA-bilayer graphene in the form $\hat{H}=\hat{H}_{0}+\hat{H}_{V}$, where $\hat{H}_{0}$ is the Hamiltonian of the system without the applied electric field potential. The Hamiltonian $\hat{H}_{V}$ takes into account the effect of the external electric field potential $V$ applied to the system (see, in Fig.~\ref{fig:Fig_1}). Without any restriction on the filling (such as the half-filling, for example) in the layers in the AA-BLG system, we can write for $\hat{H}_{0}$
\begin{eqnarray}
	\hat{H}_{0}&&=-\gamma_0\sum_{\left\langle {\bf{r}}{\bf{r}}'\right\rangle}\sum_{\ell\sigma}\left({\hat{a}}^{\dag}_{\ell\sigma}({\bf{r}}){\hat{b}}_{\ell\sigma}({\bf{r}}')+{\rm h.c.}\right)
	\nonumber\\
	&&-\gamma_1\sum_{{\bf{r}}\sigma}\left({\hat{a}}^{\dag}_{1\sigma}({\bf{r}}){\hat{a}}_{2\sigma}({\bf{r}})+{\rm h.c.}\right)
	\nonumber\\
	&&-\gamma_1\sum_{{\bf{r}}\sigma}\left({\hat{b}}^{\dag}_{1\sigma}({\bf{r}}){\hat{b}}_{2\sigma}({\bf{r}})+{\rm h.c.}\right)
	\nonumber\\
	&&-\mu\sum_{\ell=1,2}\sum_{{\bf{r}}\eta}\hat{n}_{\ell\eta}({\bf{r}})
	+U\sum_{{\bf{r}}}\sum_{\ell\eta}\hat{n}_{\ell\eta\uparrow}({\bf{r}})\hat{n}_{\ell\eta\downarrow}({\bf{r}})
	\nonumber\\
	&&+W\sum_{{\bf{r}}\sigma\sigma'}\hat{n}_{a_1\sigma}({\bf{r}})\hat{n}_{{a_2}\sigma'}({\bf{r}})
	+W\sum_{{\bf{r}}\sigma\sigma'}\hat{n}_{b_1\sigma}({\bf{r}})\hat{n}_{{b_2}\sigma'}({\bf{r}}).
	\nonumber\\
	\label{Equation_1}
\end{eqnarray}

The Hamiltonian in Eq.(\ref{Equation_1}) represents the most general form of the bilayer generalization of the usual Hubbard model. It contains the usual intralayer tight-binding part, given by the interatomic matrix elements $\gamma_0$ between the adjacent (only between ${\bf{r}}$ and ${\bf{r}}'$) atomic orbitals. The Hamiltonian, in Eq.(\ref{Equation_1}), with only $\gamma_0$ hopping elements and the short-range electron-electron interactions $U$ forms the subject of the usual Hubbard model. The generalization consists in addition of the local interlayer hopping terms $\gamma_1$ and also the local interlayer electron-electron interaction terms $W$. The first one will bring only the atomic energy shift due to the potential of the neighbouring atoms in different layers while the second one could lead to the strong modifications of the usual tight-binding results similar to the Hubbard terms $U$, given in Eq.(\ref{Equation_1}). As we will see later on in this paper, the interaction terms $W$ are responsible for the formation of the interlayer excitons. It seems that the model, given in Eq.(\ref{Equation_1}) is the must accurate and complete for the description of the complicated electronic correlations in the bilayer graphene structure, presented in Fig.~\ref{fig:Fig_1}. 

We consider here the electrons, in terms of the creation and annihilation operators ${\hat{a}}^{\dag}_{\ell\sigma}({\bf{r}}),{\hat{b}}^{\dag}_{\ell\sigma}({\bf{r}})$ and ${\hat{a}}_{\ell\sigma}({\bf{r}}),{\hat{b}}_{\ell\sigma}({\bf{r}})$ entering in the covalent bonds in the separate graphene's  layers, and attached with the atoms near the atomic sites positions $A_1$, $B_1$ (in the layer with $\ell=1$) and $A_2$, $B_2$ (in the layer with $\ell=2$). The parentheses $\left\langle ...\right\rangle$, in the summation, in the first term in Eq.(\ref{Equation_1}), denote the summation over the nearest neighbour lattice sites, and $\ell$ is indexing the layers, i.e., the value $\ell=1$ corresponds to the layer $1$ and the value $\ell=2$ denotes the layer $2$ (see, in Fig.~\ref{fig:Fig_1}). The summation index $\eta$, in Eq.(\ref{Equation_1}), indicates the type of the particles, i.e., 
\begin{eqnarray}
\footnotesize
\arraycolsep=0pt
\medmuskip = 0mu
\eta
=\left\{
\begin{array}{cc}
\displaystyle   a_{1}, b_{1}, \ \ \  $if$ \ \ \ \ell=1,
\newline\\
\newline\\
{\footnotesize
	\arraycolsep=0pt
	\medmuskip = 0mu
	\begin{array}{cc}
	\displaystyle  & a_{2}, b_{2}, \ \ \  $if$ \ \ \  \ell=2.
	\end{array}}
\end{array}\right.
\label{Equation_2}
\end{eqnarray}
Next, $\mu$ is the chemical potential in the layers and we suppose that it is initially (when the system is not exposed to the electric field) the same in different layers of the BLG. The parameter $\gamma_0$ is the intralayer hopping amplitude and the parameter $\gamma_1$ is the interlayer hopping parameter (the values, found experimentally, for those parameters are $\gamma_0\sim 3$ eV and $\gamma_1=0.257$ eV, see, in Ref.\cite{cite_52}). We suppose here the homogeneous distribution of the intralayer Coulomb interaction parameter $U$, in both layers. The parameter, $W$, in the last two terms in the Hamiltonian $\hat{H}_{0}$, denotes the interlayer Coulomb interaction potential. The total density operator of the particles $\hat{n}_{\ell\eta}({\bf{r}})$ is defined as follows  
\begin{eqnarray}
\hat{n}_{\ell\eta}({\bf{r}})=\sum_{\sigma}\hat{n}_{\ell\eta\sigma}({\bf{r}}),
\label{Equation_3}
\end{eqnarray}
where, the spin-dependent fermionic density operator $\hat{n}_{\ell\eta\sigma}({\bf{r}})$ is defined as 
\begin{eqnarray}
\hat{n}_{\ell\eta\sigma}({\bf{r}})=\hat{\eta}^{\dag}_{\ell\sigma}({\bf{r}})\hat\eta_{\ell\sigma}({\bf{r}}),
\label{Equation_4}
\end{eqnarray}
here, $\sigma$ denotes the spin-variable and takes two possible values: $\sigma=\uparrow$ or $\sigma=\downarrow$. 
Next, the interlayer interaction potential $W$, in Eq.(\ref{Equation_1}), is coupled to the product of particle density operators $\hat{n}_{\eta\sigma}\hat{n}_{\eta'\sigma}$ with $\eta$ and $\eta'$ ($\eta\neq\eta'$) defined in Eq.(\ref{Equation_3}). These are the sublattices fermionic density operators and are defined as
\begin{eqnarray}
\hat{n}_{\eta\sigma}({\bf{r}})=\hat\eta^{\dag}_{\sigma}({\bf{r}})\hat\eta_{\sigma}({\bf{r}}).
\label{Equation_5}
\end{eqnarray} 

We suppose that when applying the external potential $V$ to the system, the top layer with $\ell=2$ is connected to the wire with terminal, at the potential $+V/2$, and the lower layer to the wire with terminal at the potential $-V/2$ (see, in Fig.~\ref{fig:Fig_1}). Next, we write the expression for the second part of total Hamiltonian that describes the effects of external electric field potential $V$, coupled to the electronic density operators:
\begin{eqnarray}
	\hat{H}_{V}=\frac{V}{2}\sum_{{\bf{r}}\sigma}\left(\hat{n}_{2\sigma}({\bf{r}})-\hat{n}_{1\sigma}({\bf{r}})\right).
	\label{Equation_6}
\end{eqnarray}
The densities $\hat{n}_{2\sigma}({\bf{r}})$ and $\hat{n}_{1\sigma}({\bf{r}})$, in Eq.(\ref{Equation_6}), are total fermionic densities in the layers $\ell=1,2$. They are defined as 
\begin{eqnarray}
\hat{n}_{1\sigma}({\bf{r}})=\sum_{\eta=a_{1}b_{1}}{\hat{\eta}}^{\dag}_{\sigma}({\bf{r}}){\hat{\eta}}_{\sigma}({\bf{r}}),
\nonumber\\
\hat{n}_{2\sigma}({\bf{r}})=\sum_{\eta=a_{2}b_{2}}{\hat{\eta}}^{\dag}_{\sigma}({\bf{r}}){\hat{\eta}}_{\sigma}({\bf{r}}).
\label{Equation_7}
\end{eqnarray}
In the next section, we will develop the mean-field (MF) Hartree-Fock theory for the AA-BLG with the antiferromagnetism. 
%
\section{\label{sec:Section_3} Theoretical setup}
%
\subsection{\label{sec:Section_3_1} The AA-BLG action and functional formulation}
%
In this section, we introduce the fermionic Grassmann variables on the fermionic operator field (with the principal property of complex conjugation between the operators and Grassmann variables: 
\begin{eqnarray}
\left(\eta_{\ell}\phi_{\ell\eta}\right)^{\dag}=\bar{\phi}_{_{\ell\eta}}\eta^{\dag}_{\ell},
\label{Equation_8}
\end{eqnarray}
where $\bar{\phi}_{_{\ell\eta}}$ signifies the complex conjugation for the Grassmann field (contrary to the Hermitian conjugation, defined by the symbol $\dag$, for the operators). For our problem, we have
\begin{eqnarray}
\footnotesize
\arraycolsep=0pt
\medmuskip = 0mu
\left({\phi}_{{\ell\eta}}, \bar{\phi}_{{\ell\eta}}\right)
=\left\{
\begin{array}{cc}
\displaystyle   \left({a}_{1},\bar{a}_{1}\right), \ \ \  $if$ \ \ \ \ell=1 \ \ \ $and$ \ \ \ \eta=A_{1},
\newline\\
\newline\\
{\footnotesize
	\arraycolsep=0pt
	\medmuskip = 0mu
	\begin{array}{cc}
	\displaystyle  & \left({b}_{1},\bar{b}_{1}\right), \ \ \  $if$ \ \ \  \ell=1 \ \ \ $and$ \ \ \ \eta=B_{1},
	\newline\\
	\newline\\
	\displaystyle  & \left({a}_{2},\bar{a}_{2}\right), \ \ \  $if$ \ \ \  \ell=1 \ \ \ $and$ \ \ \ \eta=A_{2},
	\newline\\
	\newline\\
	\displaystyle  & \left({b}_{2},\bar{b}_{2}\right), \ \ \  $if$ \ \ \  \ell=1 \ \ \ $and$ \ \ \ \eta=B_{2}.
	\end{array}}
\end{array}\right.
\label{Equation_9}
\end{eqnarray}
Then, we pass from the fermionic operator representation into the Grassmann representation (see, a similar description, in Refs.\cite{cite_53, cite_54}) by associating the fermionic operators ${\hat{a}}^{\dag}_{1}, {\hat{b}}^{\dag}_{1}, {\hat{a}}^{\dag}_{2}$, and ${\hat{b}}^{\dag}_{2}$ with the Grassmann complex variables $\bar{\phi}_{{\ell\eta}}$, introduced above. We write the partition function ${\cal{Z}}$ of the AA-BLG system in the fermionic path integral formulation (see, in Ref.\cite{cite_53, cite_54})  
\begin{eqnarray}
{\cal{Z}}=\mathrm{Tr}{e^{-\beta{\hat{H}}}}=\int\left[{{\cal D}\bar{a}{\cal{D}}}a\right]\int\left[{{\cal{D}}\bar{b}{\cal{D}}b}\right]e^{-{\cal{S}}\left[\bar{a},a,\bar{b},b\right]},
\label{Equation_10}
\end{eqnarray}
where ${\cal{S}}\left[\bar{a},a,\bar{b},b\right]$ is the total fermionic action of the system in terms of Grassmann variables and is given in the imaginary time Matsubara representation. It can be expressed with the help of total Hamiltonian ${\cal{H}}(\tau)$ (which is just the total Hamiltonian $\hat{H}$, in Grassmann-Matsubara notations) as
\begin{eqnarray}
{\cal{S}}\left[\bar{a},a,\bar{b},b\right]=\int^{\beta}_{0}d\tau{{\cal{H}}(\tau)}+\sum_{\substack{\eta=a_{1}b_{1},\\ a_{2}b_{2}}}{\cal{S}}_{\rm B}\left[\bar{\eta},\eta\right].
\label{Equation_11}
\end{eqnarray}
The upper limit of integration, in the first term in right hand side in Eq.(\ref{Equation_11}), is given after the imaginary time Matsubara formalism (with $0<\tau<\beta$) \cite{cite_55}, and we have $\beta=1/T$ (here we used the convention $k_{B}=1$).  
Next, $S_{\rm B}\left[\bar{\eta},\eta\right]$, are the Berry terms \cite{cite_53, cite_54} and are given by 
\begin{eqnarray}
{\cal{S}}_{\rm B}\left[\bar{\eta},\eta\right]=\sum_{{\bf{r}}\sigma}\int^{\beta}_{0}d\tau \bar{\eta}_{\sigma}({\bf{r}}\tau)\partial_{\tau}\eta_{\sigma}({\bf{r}}\tau).
\label{Equation_12}
\end{eqnarray}
In the next sections, we will use the total fermionic action, written in Eq.(\ref{Equation_11}), to calculate the Green's functions matrices, and to derive the set of self-consistent equations, in the considered problem.  
%
\subsection{\label{sec:Section_3_2} Mean-field decoupling and order parameters}
%
We see, in Eq.(\ref{Equation_1}), that the Hamiltonian $H_{0}$ contains the non-linear density terms (biquadratic in fermionic operators $\eta_{\ell\sigma}$ and $\bar{\eta}_{\ell\sigma}$). We can linearize these terms via the Hubbard-Stratanovich transformation rules. First of all, let's provide the following notations 
\begin{eqnarray}
{\hat{n}}_{\ell\eta}({\bf{r}})={\hat{n}}_{\ell\eta\uparrow}({\bf{r}})+{\hat{n}}_{\ell\eta\downarrow}({\bf{r}}),
\nonumber\\
{\hat{p}}_{{\rm z}\ell\eta}({\bf{r}})={\hat{n}}_{\ell\eta\uparrow}({\bf{r}})-{\hat{n}}_{\ell\eta\downarrow}({\bf{r}}). 
\label{Equation_13}
\end{eqnarray}
The last quantity ${\hat{p}}_{{\rm z}\ell\eta}({\bf{r}})$ describes indeed the polarization of the electron gas density, with respect to the orientations of the spins of cinstutent particles. Therefore, the term of type ${\hat{n}}_{\ell\eta\uparrow}({\bf{r}}){\hat{n}}_{\ell\eta\downarrow}({\bf{r}})$, in the Eq.(\ref{Equation_1}), can be rewritten in more convenient form
\begin{eqnarray}
{\hat{n}}_{\ell\eta\uparrow}({\bf{r}}){\hat{n}}_{\ell\eta\downarrow}({\bf{r}})=\frac{1}{4}\left({\hat{n}}^{2}_{\ell\eta}({\bf{r}})-{\hat{p}}^{2}_{{\rm z}\ell\eta}({\bf{r}})\right),
\label{Equation_14}
\end{eqnarray}    
where the density operator ${\hat{n}}_{\ell\eta}({\bf{r}})$ describes the total electron density in the given layer ($\ell$) and for the given type of fermions ($\eta$): ${\hat{n}}_{\ell\eta}({\bf{r}})={\hat{n}}_{\ell\eta\uparrow}({\bf{r}})+{\hat{n}}_{\ell\eta\downarrow}({\bf{r}})$. 

Next, we show the decoupling procedure of the intralayer Coulomb interaction $U$-terms in the Hamiltonian, in Eq.({\ref{Equation_1}}) (written in terms of the Grassmann algebra). For the given lattice site position ${\bf{r}}$, and at the given Matsubara time $\tau$, we can write 
\begin{eqnarray}
e^{-\frac{U}{4}\sum_{{\bf{r}}}\int^{\beta}_{0}d\tau n^{2}_{\ell\eta}({\bf{r}}\tau)}&&=\int{{\cal{D}}\xi_{\ell \eta}}\exp\left[\sum_{{\bf{r}}}\int^{\beta}_{0}d\tau\left(-\frac{1}{U}\xi^{2}_{\ell\eta}({\bf{r}}\tau)\right.\right.
\nonumber\\	
&&\left.\left.+i\xi_{\ell\eta}({\bf{r}}\tau)n_{\ell\eta}({\bf{r}}\tau)\right)\right]=\int{{\cal{D}}\xi_{\ell \eta}}e^{-{\cal{S}}[\xi]},
\label{Equation_15}
\end{eqnarray}
where the action ${\cal{S}}[\xi]$ in the exponential in the right-hand side in Eq.(\ref{Equation_15}) is of the form
\begin{eqnarray}
{\cal{S}}[\xi]=\sum_{{\bf{r}}}\int^{\beta}_{0}d\tau\left(\frac{1}{U}\xi^{2}_{\ell\eta}({\bf{r}}\tau)-i\xi_{\ell\eta}({\bf{r}}\tau)n_{\ell\eta}({\bf{r}}\tau)\right).
\label{Equation_16}
\end{eqnarray}
The integral in the right-hand side, in Eq.(\ref{Equation_15}), can be calculated in the saddle-point approximation method (which involves the functional derivation of the integral with respect to the introduced decoupling field $\xi({\bf{r}}\tau)$). For the saddle-point value of the field $\xi_{\ell\eta}({\bf{r}}\tau)$, for the given sublattice $\eta$ (in the layer $\ell$), we get
\begin{eqnarray}
	\xi^{\rm s.p}_{\ell\eta}=i\frac{U}{2}\left\langle{n}_{\ell\eta}({\bf{r}}\tau)\right\rangle=i\frac{U}{2}\bar{n}_{\ell\eta}.
\label{Equation_17}
\end{eqnarray}
Here, $\left\langle {n}_{\ell\eta}({\bf{r}}\tau)\right\rangle\equiv \bar{n}_{\ell\eta}$ is the statistical average, defined with the help of the partition function ${\cal{Z}}$, in Eq.(\ref{Equation_10}), and fermionic action in Eq.(\ref{Equation_11}). We have
\begin{eqnarray}
\left\langle... \right\rangle=\frac{1}{{\cal{Z}}}\int{... e^{-{\cal{S}}\left[\bar{a},a,\bar{b},b\right]}}.	
\label{Equation_18}
\end{eqnarray}
Then, as the mean-field approximation, we put the saddle-point value $\xi^{\rm s.p}_{\ell\eta}$, obtained for the decoupling field $\xi_{\ell\eta}({\bf{r}}\tau)$, in the expression of the action ${\cal{S}}[\xi]$ and we replace the integration over the field $\xi_{\ell\eta}({\bf{r}}\tau)$ by the value of the exponential at $\xi^{\rm s.p}_{\ell\eta}$, i.e., $\int{{\cal{D}}\xi_{\ell \eta}}e^{-{\cal{S}}[\xi]}\approx e^{-{\cal{S}}\left[\xi^{\rm s.p}_{\ell\eta}\right]}$. Furthermore, we neglect the terms which give the constant contribution to the action and, finally, we get the contribution to the Hamiltonian $H$, coming from this type of decoupling procedure. It is 
\begin{eqnarray}
	\Delta{{\cal{H}}}^{(1)}_{U}({\bf{r}}\tau)=\frac{U}{2}\sum_{\substack{\eta=a_{1}b_{1},\\ a_{2}b_{2}}}{n}_{\ell\eta}({\bf{r}}\tau)\bar{n}_{\ell\eta}.
	\label{Equation_19}
\end{eqnarray}
The decoupling of the polarization term, entering in Eq.(\ref{Equation_14}), is very similar with that given in Eq.(\ref{Equation_15}). When decoupling this term, we have
\begin{eqnarray}
e^{\frac{U}{4}\sum_{{\bf{r}}}\int^{\beta}_{0}d\tau p^{2}_{z\ell\eta}({\bf{r}}\tau)}&&=\int{{\cal{D}}\zeta_{\ell \eta}}\exp\left[\sum_{{\bf{r}}}\int^{\beta}_{0}d\tau\left(-\frac{1}{U}\zeta^{2}_{\ell\eta}({\bf{r}}\tau)\right.\right.
\nonumber\\
&&\left.\left.+\zeta_{\ell \eta}({\bf{r}}\tau)p_{z\ell\eta}({\bf{r}}\tau)\right)\right].
\label{Equation_20}
\end{eqnarray}
After performing the saddle-point approximation, we get for the polarization term
\begin{eqnarray}
\zeta_{\ell \eta}^{s.p.}=\frac{U}{2}\left\langle p_{z\ell\eta}({\bf{r}}\tau)\right\rangle\equiv \Delta^{\eta}_{\rm AFM},
\label{Equation_21}
\end{eqnarray} 
where $\Delta^{\eta}_{\rm AFM}$ is the antiferromagnetic order parameter in our problem. Then, we replace the integral in Eq.(\ref{Equation_19}) by the value at the saddle-point and we obtain the contribution to the total Hamiltonian 
\begin{eqnarray}
\Delta{{\cal{H}}}^{(2)}_{U}({\bf{r}}\tau)=\sum_{\substack{\eta=a_{1}b_{1},\\ a_{2}b_{2}}}{n}_{\ell\eta}({\bf{r}}\tau)\Delta^{\eta_{\ell}}_{\rm AFM}.
\label{Equation_22}
\end{eqnarray}
The explicit form of the antiferromagnetic order parameter $\Delta^{\eta}_{\rm AFM}$ is
\begin{eqnarray}
&\Delta^{\eta_{\ell}}_{\rm AFM}=\frac{U}{2}\bar{p}_{z\ell\eta}=
\nonumber\\
&=\frac{U}{2}\left(\left\langle \bar{\eta}_{\ell\uparrow}({\bf{r}}\tau)\eta_{\ell\uparrow}({\bf{r}}\tau)\right\rangle-\left\langle \bar{\eta}_{\ell\downarrow}({\bf{r}}\tau)\eta_{\ell\downarrow}({\bf{r}}\tau)\right\rangle\right).
\label{Equation_23}
\end{eqnarray}
We will suppose here the antiferromagnetic spin ordering in the layers (at the adjacent atomic sites positions, in the layers), and also between the layers, in the AA type stacked BLG construction. This type of antiferromagnetic ordering is called the spin $\pi$-phase or $G$-type ordering, in the literature, \cite{cite_2, cite_27, cite_28, cite_29, cite_44, cite_45}.  This means that the electrons which participate in the formation of the strong covalent bonds between the nearest neighbour atoms on the lattice have strongly opposite spin orientations. Besides, we suppose also the opposite spin orientations for the remaining electrons (which do not participate in the formation of the covalent bonds, typical for graphene materials) which are situated along the line, passing perpendicularly through the layers, in the AA-BLG structure. For the single-particle electron concentrations, with the given spin directions at the adjacent lattice sites, we have, accordingly, the following relations (typical for the $G$-type or spin-$\pi$ ordering) 
\begin{eqnarray}
\bar{n}_{a_{1}\uparrow}=\bar{n}_{b_{1}\downarrow},
\nonumber\\
\bar{n}_{a_{1}\downarrow}=\bar{n}_{b_{1}\uparrow},
\nonumber\\
\bar{n}_{a_{2}\uparrow}=\bar{n}_{b_{2}\downarrow},
\nonumber\\
\bar{n}_{a_{2}\downarrow}=\bar{n}_{b_{2}\uparrow}.
\label{Equation_24}
\end{eqnarray}
Therefore, the antiferromagnetic order parameter $\Delta^{\rm \eta}_{\rm AFM}$ takes the opposite values at the nearest neighbor lattice sites $\eta$ in the give layer. Across the interlayer stacking direction this property remains the same. 
After having in mind the antiferromagnetic $G$-type spin ordering, between the layers in the AA-BLG construction we have  
\begin{eqnarray}
\Delta^{\rm a_{1}}_{\rm AFM}=-\Delta^{\rm b_{1}}_{\rm AFM}=-\Delta^{\rm a_{2}}_{\rm AFM}=\Delta^{\rm b_{2}}_{\rm AFM}.
\label{Equation_25}
\end{eqnarray}
Thus, by putting $\Delta^{a_{1}}_{\rm AFM}\equiv \Delta_{\rm AFM}$ we can write 
\begin{eqnarray}
\Delta^{a_{2}}_{\rm AFM}=\Delta^{b_{1}}_{\rm AFM}=-\Delta^{a_{1}}_{\rm AFM}=-\Delta_{\rm AFM}.
\label{Equation_26}
\end{eqnarray}
and $\Delta^{b_{2}}_{\rm AFM}\equiv\Delta_{\rm AFM}$.
The physical sense of this parameter lays in the charge density imbalance for different spin orientations mediated by the intralayer  Coulomb coupling parameter $U$. In turn, the excitonic order parameter appears after decoupling of the interlayer Coulomb interaction terms (see, the last two terms, in Eq.(\ref{Equation_1})). We discuss shortly here the decoupling procedure of those terms. Indeed, we have for those terms the following relations (in the operator notations)  
\begin{eqnarray}
&W\sum_{{\bf{r}}\sigma\sigma'}{\hat{n}}_{a_1\sigma}({\bf{r}}){\hat{n}}_{{a_2}\sigma'}({\bf{r}})=2W\sum_{{\bf{r}}\sigma}{\hat{n}}_{{a_2}\sigma'}({\bf{r}})
\nonumber\\
&-W\sum_{{\bf{r}}\sigma\sigma'}{\hat{\Delta}}^{\dag(a)}_{\sigma'\sigma}({\bf{r}}){\hat{\Delta}}^{(a)}_{\sigma'\sigma}({\bf{r}})=
\nonumber\\
&=2W\sum_{{\bf{r}}\sigma}{\hat{n}}_{{a_2}\sigma'}({\bf{r}})-W\sum_{{\bf{r}}\sigma\sigma'}|{\hat{\Delta}}^{(a)}_{\sigma'\sigma}({\bf{r}})|^{2}
\nonumber\\
&W\sum_{{\bf{r}}\sigma\sigma'}{\hat{n}}_{b_1\sigma}({\bf{r}}){\hat{n}}_{{b_2}\sigma'}({\bf{r}})=2W\sum_{{\bf{r}}\sigma}{\hat{n}}_{{b_2}\sigma'}({\bf{r}})
\nonumber\\
&-W\sum_{{\bf{r}}\sigma\sigma'}{\hat{\Delta}}^{\dag(b)}_{\sigma'\sigma}({\bf{r}}){\hat{\Delta}}^{(b)}_{\sigma'\sigma}({\bf{r}})=
\nonumber\\
&=2W\sum_{{\bf{r}}\sigma}{\hat{n}}_{{b_2}\sigma'}({\bf{r}})-W\sum_{{\bf{r}}\sigma\sigma'}|{\hat{\Delta}}^{(b)}_{\sigma'\sigma}({\bf{r}})|^{2}.
\label{Equation_27}
\end{eqnarray}
First terms, in both equations Eq.(\ref{Equation_26}), contribute to the chemical potentials (as energy shifts of  the chemical potential $\mu$ in Eq.(\ref{Equation_1})). The second terms, in the right hand sides in equations, in Eq.(\ref{Equation_26}), are written with the help of the parameters ${\hat{\Delta}}^{(a)}_{\sigma'\sigma}({\bf{r}})$, ${\hat{\Delta}}^{(b)}_{\sigma'\sigma}({\bf{r}})$ and their Hermitian conjugates, as 
\begin{eqnarray}
{\hat{\Delta}}^{(a)}_{\sigma'\sigma}({\bf{r}})={\hat{a}}^{\dag}_{1\sigma'}({\bf{r}}){\hat{a}}_{2\sigma'}({\bf{r}}),
\nonumber\\
{\hat{\Delta}}^{\dag(a)}_{\sigma'\sigma}({\bf{r}})={\hat{a}}^{+}_{2\sigma'}({\bf{r}}){\hat{a}}_{1\sigma'}({\bf{r}}).
\nonumber\\
{\hat{\Delta}}^{(b)}_{\sigma'\sigma}({\bf{r}})={\hat{b}}^{\dag}_{1\sigma'}({\bf{r}}){\hat{b}}_{2\sigma'}({\bf{r}}),
\nonumber\\
{\hat{\Delta}}^{\dag(b)}_{\sigma'\sigma}({\bf{r}})={\hat{b}}^{+}_{2\sigma'}({\bf{r}}){\hat{b}}_{1\sigma'}({\bf{r}}).
\label{Equation_28}
\end{eqnarray} 
The new parameters ${\hat{\Delta}}^{(\eta)}_{\sigma'\sigma}({\bf{r}})$ (with $\eta=a,b$) are indeed the subjects of complex matrices $2\times 2$, if we take into account all spin orientations $\sigma=\uparrow, \downarrow$ and $\sigma'=\uparrow, \downarrow$. Thus, we have, in general
\begin{eqnarray}
\mathbf{{\hat{\Delta}}}^{(\eta)}({\bf{r}})=
\left(\begin{matrix}
{\hat{\Delta}}^{(\eta)}_{\uparrow\uparrow}({\bf{r}}) & {\hat{\Delta}}^{(\eta)}_{\uparrow\downarrow}({\bf{r}}) & \\
{\hat{\Delta}}^{(\eta)}_{\downarrow\uparrow}({\bf{r}}) & {\hat{\Delta}}^{(\eta)}_{\downarrow\downarrow}({\bf{r}}).&  
\end{matrix}\right).
\label{Equation_29}
\end{eqnarray}
Here, the non-diagonal elements in Eq.(\ref{Equation_28}) vanishes, due to the symmetry of the total action of the system, in Eq.(\ref{Equation_43}), and we calculate the diagonal terms of these matrices. We denote them with single spin indices $\sigma$ as ${\hat{\Delta}}^{(\eta)}_{\sigma\sigma}({\bf{r}})\equiv{\hat{\Delta}}^{(\eta)}_{\sigma}({\bf{r}})$. The decoupling of the last terms (here, we are redialing the Grassmann variables, introduced at the beginning of this section) in Eq.(\ref{Equation_26}) could be also done within the path integral formulation. We have
\begin{eqnarray}
&&e^{W\sum_{{\bf{r}}}\int^{\beta}_{0}d\tau|\Delta^{(\eta)}_{\sigma}({\bf{r}}\tau)|^{2}}=
\nonumber\\
&&=\int{\left[{\cal{D}}\bar{\Xi}{\cal{D}}{\Xi}\right]}\exp\left[\sum_{{\bf{r}}}\int^{\beta}_{0}d\tau\left(-\frac{1}{W}|\Xi^{(\eta)}_{\sigma}({\bf{r}}\tau)|^{2}\right.\right.
\nonumber\\
&&\left.\left.+\bar{\Xi}^{(\eta)}_{\sigma}({\bf{r}}\tau)\Delta^{(\eta)}_{\sigma}({\bf{r}}\tau)+{\Xi}^{(\eta)}_{\sigma}({\bf{r}}\tau)\bar{\Delta}^{(\eta)}_{\sigma}({\bf{r}}\tau)\right)\right].
\label{Equation_30}
\end{eqnarray}
Next, we perform the saddle-point approximation for the integral in the right-hand side in Eq.(\ref{Equation_29}) and we get the saddle-point value for the external source parameters ${\Xi}^{(\eta)}_{\sigma}({\bf{r}}\tau)$ and $\bar{\Xi}^{(\eta)}_{\sigma}({\bf{r}}\tau)$. We get 
\begin{eqnarray}
{\Xi}^{(\eta)s.p.}_{\sigma}=W\left\langle \Delta^{(\eta)}_{\sigma}({\bf{r}}\tau)\right\rangle=
\nonumber\\
=W\left\langle \bar{\eta}_{1\sigma'}({\bf{r}}\tau)\eta_{2\sigma'}({\bf{r}}\tau)\right\rangle\equiv\Delta^{\left(\eta\right)}_{\rm exc{\sigma}}.
\label{Equation_31}
\end{eqnarray}
The saddle-point values of the parameters ${\Xi}^{(\eta)}_{\sigma}({\bf{r}}\tau)$ represents, in fact, the excitonic pairing parameters between the same $\eta$-type fermions situated in different layers in the AA-BLG system. In the rest of the paper we call those saddle-point values as the excitonic order parameters $\Delta^{{\rm \left(\eta\right)}}_{\rm exc{\sigma}}$ (with $\eta=a,b$). The contribution to the total Hamiltonian, after decoupling of $W$-terms, presented here, will be 
\begin{eqnarray}
\Delta{{\cal{H}}}_{W}({\bf{r}}\tau)=\sum_{\substack{\eta=a_{1}b_{1},\\ a_{2}b_{2}}}\sum_{\sigma}\Delta^{\left(\eta\right)}_{\rm exc{\sigma}}\Delta^{(\eta)}_{\sigma}({\bf{r}}\tau).
\label{Equation_32}
\end{eqnarray}
In the next section, we will write the total MF Hamiltonian after decoupling procedure and we will give the total MF action in the AA-BLG.
%
\subsection{\label{sec:Section_3_3} MF effective Hamiltonian}
%
Here, we write the total MF Hamiltonian of the system ${\cal{H}}_{\rm MF}$ obtained after the decoupling procedures, described in the Section \ref{sec:Section_3_2}. It reads as 
\begin{eqnarray} 
&&{\cal{H}}_{\rm MF}(\tau)={\cal{H}}_{\gamma_0}\left(\tau\right)+{\cal{H}}_{\gamma_1}\left(\tau\right)+{\cal{H}}_{\mu}\left(\tau\right)+\Delta{{\cal{H}}}^{(1)}_{U}(\tau)
\nonumber\\
&&+\Delta{{\cal{H}}}^{(2)}_{U}(\tau)+\Delta{{\cal{H}}}_{W}(\tau).
\label{Equation_33}
\end{eqnarray}
Here, the first three terms are Grassmann versions of the corresponding terms, in the Hamiltonian ${\cal{H}}$ with the operator notations (see, in Eq.(\ref{Equation_1})). They are given as 
\begin{eqnarray} 
&&{\cal{H}}_{\gamma_0}\left(\tau\right)=-\gamma_0\sum_{\left\langle {\bf{r}}{\bf{r}}'\right\rangle}\sum_{\ell\sigma}\left({\bar{a}}_{\ell\sigma}({\bf{r}}\tau){b}_{\ell\sigma}({\bf{r}}'\tau)+{\rm h.c.}\right),
\nonumber\\
&&{\cal{H}}_{\gamma_1}\left(\tau\right)=-\gamma_1\sum_{{\bf{r}}\sigma}\left({\bar{a}}_{1\sigma}({\bf{r}}\tau){{a}}_{2\sigma}({\bf{r}}\tau)+{\rm h.c.}\right)
\nonumber\\
&&-\gamma_1\sum_{{\bf{r}}\sigma}\left({\bar{b}}_{1\sigma}({\bf{r}}\tau){b}_{2\sigma}({\bf{r}}\tau)+{\rm h.c.}\right)
\label{Equation_34}
\end{eqnarray}
and
\begin{eqnarray}
&&{\cal{H}}_{\mu}\left(\tau\right)=-\mu\sum_{\ell=1,2}\sum_{{\bf{r}}}\sum_{\eta=\substack{a_{1}b_{1}, \\ a_{2}b_{2}}}{n}_{\ell}({\bf{r}}\tau).
\label{Equation_35}
\end{eqnarray}
The last three terms in the Hamiltonian, in Eq.(\ref{Equation_32}), represent the ${\bf{r}}$-summed contributions of the partial Hamiltonians, given in Eqs.(\ref{Equation_18}), (\ref{Equation_21}) and (\ref{Equation_31}). 
\begin{eqnarray}
&&\Delta{{\cal{H}}}^{(1)}_{U}(\tau)=\sum_{{\bf{r}}}\Delta{{\cal{H}}}^{(1)}_{U}({\bf{r}}\tau),
\nonumber\\
&&\Delta{{\cal{H}}}^{(2)}_{U}(\tau)=\sum_{{\bf{r}}}\Delta{{\cal{H}}}^{(2)}_{U}({\bf{r}}\tau),
\nonumber\\
&&\Delta{{\cal{H}}}_{W}(\tau)=\sum_{{\bf{r}}}\Delta{{\cal{H}}}_{W}({\bf{r}}\tau).
\label{Equation_36}
\end{eqnarray}
Furthermore, we pass into the reciprocal space representation $\left({\bf{k}},\nu_{n}\right)$  (where ${\bf{k}}$ is the wave vector in the reciprocal space, conjugated to the vector ${\bf{r}}$ in the real space) for the fermionic variables $\eta$ and $\bar{\eta}$. This could be done by Fourier transformation
\begin{eqnarray}
\eta_{\ell\sigma}({\bf{r}}\tau)=\frac{1}{\beta{N}}\sum_{{\bf{k}}\nu_{n}}\eta_{\ell\sigma}({\bf{k}}\nu_{n})e^{i({\bf{k}}{\bf{r}}-\nu_{n}\tau)}, 
\label{Equation_37}
\end{eqnarray}
where $N$ is the total number of wave vectors $|{\bf{k}}|$ in the region in ${\bf{k}}$-space, corresponding to the first Brillouin zone. The frequencies $\nu_{n}$ are fermionic Matsubara frequencies $\nu_{n}=\pi{T}(2n+1)$ \cite{cite_55}, (where $n=0, \pm1, \pm2, \pm3, ...$). First, we separate the total MF Hamiltonian ${\cal{H}}_{\rm MF}$ into two parts which correspond to two different spin directions $\sigma=\uparrow$ and $\sigma=\downarrow$. We write ${\cal{H}}_{\rm MF}={\cal{H}}_{\rm MF\uparrow}+{\cal{H}}_{\rm MF\downarrow}$. Hereafter, we present the forms of the individual Hamiltonians ${\cal{H}}_{\rm MF\uparrow}$ and ${\cal{H}}_{\rm MF\downarrow}$, after the MF decoupling procedure. Particularly, for the total Hamiltonian ${\cal{H}}_{\rm MF\sigma}$, corresponding to spin direction $\sigma$, we get
\begin{widetext}
\small{\begin{eqnarray}
&&{\cal{H}}_{\rm MF\sigma}={\cal{H}}_{\gamma_0\sigma}+{\cal{H}}_{\gamma_1\sigma}-\frac{\mu_{1}+(-1)^{n_{\sigma}}\Delta_{\rm AFM}}{\beta{N}}\sum_{{\bf{k}}\nu_{n}}\bar{a}_{1\sigma}({\bf{k}}\nu_{n})a_{1\sigma}({\bf{k}}\nu_{n})-\frac{\mu_{2}-(-1)^{n_{\sigma}}\Delta_{\rm AFM}-2W}{\beta{N}}\sum_{{\bf{k}}\nu_{n}}\bar{a}_{2\sigma}({\bf{k}}\nu_{n})a_{2\sigma}({\bf{k}}\nu_{n})
\nonumber\\
&&-\frac{\mu_{1}-(-1)^{n_{\sigma}}\Delta_{\rm AFM}}{\beta{N}}\sum_{{\bf{k}}\nu_{n}}\bar{b}_{1\sigma}({\bf{k}}\nu_{n})b_{1\sigma}({\bf{k}}\nu_{n})-\frac{\mu_{2}+(-1)^{n_{\sigma}}\Delta_{\rm AFM}-2W}{\beta{N}}\sum_{{\bf{k}}\nu_{n}}\bar{b}_{2\sigma}({\bf{k}}\nu_{n})b_{2\sigma}({\bf{k}}\nu_{n})-\frac{\Delta^{\left(a\right)}_{\rm exc\sigma}}{\beta{N}}\sum_{{\bf{k}}\nu_{n}}\bar{a}_{1\sigma}({\bf{k}}\nu_{n})a_{2\sigma}({\bf{k}}\nu_{n})
\nonumber\\
&&-\frac{{\Delta}^{\ast\left(a\right)}_{\rm exc\sigma}}{\beta{N}}\sum_{{\bf{k}}\nu_{n}}\bar{a}_{2\sigma}({\bf{k}}\nu_{n})a_{1\sigma}({\bf{k}}\nu_{n})-\frac{\Delta^{\left(b\right)}_{\rm exc\sigma}}{\beta{N}}\sum_{{\bf{k}}\nu_{n}}\bar{b}_{1\sigma}({\bf{k}}\nu_{n})b_{2\sigma}({\bf{k}}\nu_{n})
-\frac{{\Delta}^{\ast\left(b\right)}_{\rm exc\sigma}}{\beta{N}}\sum_{{\bf{k}}\nu_{n}}\bar{b}_{2\sigma}({\bf{k}}\nu_{n})b_{1\sigma}({\bf{k}}\nu_{n}).
\label{Equation_38}
\end{eqnarray}}
\end{widetext}
The spin variable $\sigma$ can take two directions in our problem: $\sigma=\uparrow$ or $\sigma=\downarrow$. The number $n_{\sigma}$, in Eq.(\ref{Equation_37}), takes the values
\begin{eqnarray}
\footnotesize
n_{\sigma}=
\left\{
\begin{array}{cc}
	\displaystyle  & {\rm even}, \ \ \  $if$ \ \ \ \sigma=\uparrow,
	\newline\\
	\newline\\
	\displaystyle  & {\rm odd},  \ \ \  $if$ \ \ \ \sigma=\downarrow.
\end{array}\right.
\label{Equation_39}
\end{eqnarray}
We have introduced in Eq.(\ref{Equation_37}) the new effective chemical potentials $\mu_{1}$ and $\mu_{2}$ which emerge in the problem as 
\begin{eqnarray}
\mu_{i}=\mu+(-1)^{i+1}\frac{V}{2}-\frac{U}{2}\bar{n}_{a_{i}},
\label{Equation_40}
\end{eqnarray}
where $i=1,2$. When writing the expression in Eq.(\ref{Equation_39}) we have supposed the equal average electron concentrations at the nearest neighbour lattice site positions, in the layers $\ell=1,2$, i.e., $\bar{n}_{a_{1}}=\bar{n}_{b_{1}}$ and $\bar{n}_{a_{2}}=\bar{n}_{b_{2}}$. The fermionic Berry terms, figuring in Eqs.(\ref{Equation_11}) and (\ref{Equation_12}), are given in the ${\bf{k}}$-space as
\begin{eqnarray}
{\cal{S}}_{B}\left[\bar{\eta},\eta\right]=\frac{1}{\beta{N}}\sum_{\sigma}\sum_{{\bf{k}}\nu_{n}}\bar{\eta}_{\sigma}({\bf{k}}\nu_{n})(-i\nu_{n})\eta_{\sigma}({\bf{k}}\nu_{n}).
\label{Equation_41}
\end{eqnarray}
We see in Eq.(\ref{Equation_37}) that the form of the Hamiltonian ${\cal{H}}_{\downarrow}$ is different from the form of ${\cal{H}}_{\uparrow}$, and the difference is attributed to the sign of the antiferromagnetic order parameter $\Delta_{\rm AFM}$. One of the consequences of this observation is that the excitonic order parameter $\Delta^{\left(\eta\right)}_{{\rm exc}\sigma}$ is not spin-symmetric, and we need to derive two self-consistent equations for the order parameters $\Delta^{{\rm \left(\eta\right)}}_{\rm exc{\uparrow}}$ and $\Delta^{{\rm \left(\eta\right)}}_{\rm exc{\downarrow}}$, separately. For this, we introduce here the Nambu-Gorkov spinors ${\psi}_{\sigma}({\bf{k}}\nu_{n})$ for the considered AA-BLG system. They are defined as 
\begin{eqnarray} 
	{\psi}_{\sigma}({\bf{k}}\nu_{n})=\left(
	\begin{array}{crrrr}
		a_{1\sigma}({\bf{k}}\nu_{n})\\\\
		b_{1\sigma}({\bf{k}}\nu_{n}) \\\\
		a_{2\sigma}({\bf{k}}\nu_{n}) \\\\
		b_{2\sigma}({\bf{k}}\nu_{n}) \\\\
	\end{array}
	\right).
	\label{Equation_42}
\end{eqnarray}
The form of complex conjugate field is obvious
\begin{eqnarray} 
\bar{\psi}_{\sigma}({\bf{k}}\nu_{n})=\left(	\bar{a}_{1\sigma}({\bf{k}}\nu_{n}),\bar{b}_{1\sigma}({\bf{k}}\nu_{n}),\bar{a}_{2\sigma}({\bf{k}}\nu_{n}),\bar{b}_{2\sigma}({\bf{k}}\nu_{n})\right).
\nonumber\\
\label{Equation_43}
\end{eqnarray}
Furthermore, we write the total effective fermionic action using the notations, introduced in Eqs.(\ref{Equation_41}) and (\ref{Equation_42}). It reads as
%
%
\begin{eqnarray}
{\cal{S}}_{\rm eff}\left[\bar{\psi}, \psi\right]=\frac{1}{\beta{N}}\sum_{{\bf{k}}\nu_{n}\sigma}\bar{\psi}_{\sigma}({\bf{k}}\nu_{n}){\cal{G}}^{-1}_{\sigma}({\bf{k}}\nu_{n})\psi_{\sigma}({\bf{k}}\nu_{n}),
\label{Equation_44}
\end{eqnarray}
where the expression for the inverse Green's function matrix is

\begin{widetext}
	
\begin{eqnarray}
	{\cal{G}}^{-1}_{\sigma}({\bf{k}}\nu_{n})=
		\left(\begin{matrix}
			-i\nu_{n}-\mu_{1}+\Delta_{\rm AFM} & \tilde{\gamma}_{0{\bf{k}}} & -{\gamma}_1-\Delta^{a}_{{\rm exc}\sigma} & 0\\
			\tilde{\gamma}^{\ast}_{0{\bf{k}}} & -i\nu_{n}-\mu_{1}-\Delta_{\rm AFM} & 0 &  -{\gamma}_1-\Delta^{b}_{{\rm exc}\sigma}\\
			-{\gamma}_1-{\Delta}^{\ast{a}}_{{\rm exc}\sigma} & 0 & -i\nu_{n}-\mu_{1}+\Delta_{\rm AFM}+2W & \tilde{\gamma}_{0{\bf{k}}} \\
			0 & -{\gamma}_1-{\Delta}^{\ast{b}}_{{\rm exc}\sigma} & \tilde{\gamma}^{\ast}_{0{\bf{k}}} & -i\nu_{n}-\mu_{1}-\Delta_{\rm AFM}+2W 
		\end{matrix}\right).
		\nonumber\\
	\label{Equation_45}
\end{eqnarray}
\end{widetext}
The diagonal elements of the matrix, in Eq.(\ref{Equation_44}), describe the spectrum in the problem.
As calculations show here, the AA-BLG electronic system, with the antiferromagnetic ordering, is not symmetric with respect to the change of the direction of spin variable $\sigma$: $\uparrow \rightarrow \downarrow$.
Next, the ${\bf{k}}$-dependent parameters $\tilde{\gamma}_{0{\bf{k}}}$ are the Fourier transforms coming from the intralayer hopping terms and they are defined here as
\begin{eqnarray}
\tilde{\gamma}_{0{\bf{k}}}=\gamma_0\sum_{\bm{\mathit{\delta}}}e^{-i{{\bf{k}}\bm{\mathit{\delta}}}}.
\label{Equation_46}
\end{eqnarray}
The vectors $\bm{\mathit{\delta}}$ are the nearest neighbor vectors in different layers $\ell=1,2$. The components of $\bm{\mathit{\delta}}$ are the same for the layers with $\ell=1,2$ and are given by 
\begin{eqnarray}
\footnotesize
\bm{\mathit{\delta}}=
\left\{
\begin{array}{cc}
\displaystyle  & \bm{\mathit{\delta}}_{1}=\left({a}/{2\sqrt{3}},a/2\right),
\newline\\
\newline\\
\displaystyle  & \bm{\mathit{\delta}}_{2}=\left({a}/{2\sqrt{3}},-a/2\right),
\newline\\
\newline\\
\displaystyle  &
\bm{\mathit{\delta}}_{3}=\left(-a/\sqrt{3},0\right)
\end{array}\right.
\label{Equation_47}
\end{eqnarray}
where $a=\sqrt{3}a_{0}$ is the sublattice constant, while $a_{0}$ is the carbon-carbon length in the graphene sheets (with $a_{0}=1.42\AA$. 

It is important to notice that the inverse Green's function matrix ${\cal{G}}^{-1}_{\downarrow}({\bf{k}}\nu_{n})$, for the direction of the spin $\sigma=\downarrow$, is different from ${\cal{G}}^{-1}_{\uparrow}({\bf{k}}\nu_{n})$, i.e., 
\begin{eqnarray}
{\cal{G}}^{-1}_{\downarrow}({\bf{k}}\nu_{n})\neq{\cal{G}}^{-1}_{\uparrow}({\bf{k}}\nu_{n}).
\label{Equation_48}
\end{eqnarray}
It is particularly interesting to notice again that there are changes in the sign of the order parameter $\Delta_{\rm AFM}$ when reversing the spin direction. Thus, we have: $\Delta_{\rm AFM}\rightarrow-\Delta_{\rm AFM}$ when changing $\sigma:\uparrow \rightarrow\downarrow$. Therefore, we conclude here that spin-symmetry (and the behaviour of the AA-BLG system) in the considered problem is strongly affected by the existence of the  antiferromagnetism in the AA-BLG.

Concerning the excitonic pairing gap parameter $\Delta^{\eta}_{{\rm exc}\sigma}$ with $\eta=a, b$, we have the same homogeneous values for the all sublattice sites positions: $\Delta^{a}_{{\rm exc}\sigma}=\Delta^{b}_{{\rm exc}\sigma}\equiv \Delta_{\rm exc \sigma}$. It is important also to remark here, that, although, ${{\cal{G}}^{-1}_{\uparrow}({\bf{k}}\nu_{n})}\neq{{\cal{G}}^{-1}_{\downarrow}({\bf{k}}\nu_{n})}$, the determinants of those matrices are the same: 
\begin{eqnarray}
\det{{\cal{G}}^{-1}_{\uparrow}({\bf{k}}\nu_{n})}=\det{{\cal{G}}^{-1}_{\downarrow}({\bf{k}}\nu_{n})}\equiv \det{{\cal{G}}^{-1}({\bf{k}}\nu_{n})}. 
\label{Equation_49}
\end{eqnarray}
The principal consequence of this last artefact is that the energy excitation spectra, for both spin orientations $\sigma=\uparrow$ and $\sigma=\downarrow$, are the same, i.e., $\varepsilon_{i{\bf{k}}\uparrow}=\varepsilon_{i{\bf{k}}\downarrow}\equiv\varepsilon_{i{\bf{k}}}$, for all values of the energy branches: $i=1,...,4$. We will show in this paper that the excitonic gap parameter depends on the spin orientation, i.e., $\Delta_{\rm exc\uparrow}\neq\Delta_{\rm exc\downarrow}$ and we will calculate them numerically. The single-particle excitation spectrum is defined with the equation $\det{{\cal{G}}^{-1}({\bf{k}}\nu_{n})}=0$. We have, for both $\sigma$
\begin{eqnarray}
\prod^{4}_{i=1}(-i\nu_{n}-\varepsilon_{i{\bf{k}}}))=0.
\label{Equation_50}
\end{eqnarray}
The energy parameters $\varepsilon_{i{\bf{k}}}$ (with $i=1,...,4$) are given by the following mathematical expressions:
\begin{eqnarray}
\varepsilon_{m{\bf{k}}}=\frac{1}{2}\left[-a-(-1)^{m+1}\sqrt{b-2\left(c({{\bf{k}}})+\sqrt{d}\right)}\right],	
\nonumber\\
\varepsilon_{n{\bf{k}}}=\frac{1}{2}\left[-a-(-1)^{n+1}\sqrt{b-2\left(c({{\bf{k}}})-\sqrt{d}\right)}\right].
\label{Equation_51}
\end{eqnarray}
Here, we have $m=1,2$ and $n=3,4$.
The parameters $a,b,c({{\bf{k}}}),d$, entering in Eq.(\ref{Equation_50}), are defined as
\begin{eqnarray}
&&a=2W-\mu_{1}-\mu_{2},
\nonumber\\
&&b=a^{2},
\nonumber\\
&&c({{\bf{k}}})=-2|\tilde{\gamma}_{0{\bf{k}}}|^{2}-2\gamma_{1}\left(\Delta_{{\rm exc}\uparrow}+\Delta_{{\rm exc}\downarrow}+\gamma_1\right),
\nonumber\\
&&-\left(\Delta^{2}_{{\rm exc}\uparrow}+\Delta^{2}_{{\rm exc}\downarrow}\right)-2\left(\Delta^{2}_{\rm AFM}-4W\mu_{1}+2\mu_{1}\mu_{2}\right),
\nonumber\\
&&d=A\Delta^{2}_{\rm AFM}+B\Delta_{\rm AFM}+C.
\label{Equation_52}
\end{eqnarray} 
The coefficients $A$, $B$ and $C$, in the last expression, for the parameter $d$, are defined in the following form
\begin{eqnarray}
&&A=4\left(2W+\mu_{1}-\mu_{2}\right)^{2},
\nonumber\\
&&B=4\left(\Delta_{{\rm exc}\uparrow}-\Delta_{{\rm exc}\downarrow}\right)\left(2\gamma_1+\Delta_{{\rm exc}\uparrow}+\Delta_{{\rm exc}\downarrow}\right)\times
\nonumber\\
&&\times\left(2W+\mu_{1}-\mu_{2}\right),
\nonumber\\
&&C=\left(2\gamma_{1}+\Delta_{{\rm exc}\uparrow}+\Delta_{{\rm exc}\downarrow}\right)^{2}+\left(2W+\mu_{1}-\mu_{2}\right)^{2}.
\label{Equation_53}
\end{eqnarray} 
We will write the effective chemical potentials $\mu_{1}$ and $\mu_{2}$, given in Eq.(\ref{Equation_39}), in more computational form, dealing with the filling coefficient $\kappa$ and the average total densities difference (between the layers) $\delta_{n}$. For this, we remark first that
\begin{eqnarray}
\bar{n}_{a_{1}}+\bar{n}_{a_{2}}=\frac{1}{\kappa},
\nonumber\\
\bar{n}_{a_{2}}-\bar{n}_{a_{1}}=\frac{\delta \bar{n}}{4},
\label{Equation_54}
\end{eqnarray} 
where the parameter $\kappa$, figuring in the first equation, is the filling factor, the inverse of which indicates the average total number of particles on the sublattices $A_{1}$ and $A_{2}$. The value $\kappa=1$ could correspond to the scenario of the electron-hole type $AA$-BLG system and the regime with $\kappa=0.5$ corresponds to the half-filling regime, usually treated in the literature (remember here that the inverse of the coefficient $\kappa$ gives the exact filling factor $\kappa^{-1}$ in the $AA$ system). The parameter $\delta \bar{n}$ is defined as $\delta \bar{n}=\bar{n}_{2}-\bar{n}_{1}$. The second of equations in Eq.(\ref{Equation_53}) could be obtained from the definitions of the partial average charge densities in different layers, given in Eq.(\ref{Equation_23}). Indeed, for the total average charge densities in the layers $1$ and $2$, we can write  the following relations
\begin{eqnarray}
\bar{n}_{a_{1}}+\bar{n}_{b_{1}}=2\bar{n}_{a_{1}}=\bar{n}_{1},
\nonumber\\
\bar{n}_{a_{2}}+\bar{n}_{b_{2}}=2\bar{n}_{a_{2}}=\bar{n}_{2}.
\label{Equation_55}
\end{eqnarray} 
Next, after subtracting from the second of equations, in Eq.(\ref{Equation_53}), the first one, we get:
\begin{eqnarray}
\bar{n}_{2}-\bar{n}_{1}=\bar{n}_{a_{2}}+\bar{n}_{b_{2}}-\bar{n}_{a_{1}}-\bar{n}_{b_{1}}=2\left(\bar{n}_{a_{2}}-\bar{n}_{a_{1}}\right).
\label{Equation_56}
\end{eqnarray}
On the other hand, the difference between the intralayer average charge density operators $\bar{n}_{2}$ and $\bar{n}_{1}$ gives us the unknown parameter $\delta{\bar{n}}$. Then we get $2\left(\bar{n}_{a_{2}}-\bar{n}_{a_{1}}\right)=\delta \bar{n}$ which is exactly the second equation, in Eq.(\ref{Equation_53}).
Note, that for the value $\kappa=0.5$ we have the limit of the half-filling, widely considered in the literature (with the value ${\kappa}^{-1}=2$ corresponding to the occupation of only one particle per lattice site in different layers). To understand well the regime away from the half-filling (or the partial filling), in this case, it is sufficient to consider the spins of fermionic particles which are not strongly localized along the $z$-direction (we should especially underline that we don't mean here the fluctuating spins but, rather, the localized spins, with directions other than $z$). The number $\delta{\bar{n}}$, which should be calculated numerically (see, the next subsection, in this section), signifies the charge density imbalance between the layers in the AA-BLG structure. Also, it is worth to mention here that the parameter $\delta{\bar{n}}$ appears after applying the external electric field potential $V$ to the system and is related to the population inversion between the layers in the AA-BLG. It is interesting to write the forms of the effective chemical potentials $\mu_{1}$ and $\mu_{2}$ in terms of the charge density difference function $\delta\bar{n}$ and filling coefficient $\kappa$. We have 
\begin{eqnarray}
\mu_{i}=\mu+(-1)^{i+1}\frac{V}{2}-\frac{U}{4}\left(\frac{1}{\kappa}-(-1)^{i+1}\frac{{\delta\bar{n}}}{2}\right),
\label{Equation_57}
\end{eqnarray} 
where $i=1,2$.
In the case of exact half-filling in the layers, we have, in addition, also the following identities $\bar{n}_{a_{2}\uparrow}=\bar{n}_{a_{1}\downarrow}$ and $\bar{n}_{a_{1}\uparrow}=\bar{n}_{a_{2}\downarrow}$, which means the absence of the interlayer antiferromagnetic order parameter (mediated by the local interlayer interactions) in the $G$-type antiferromagnetic AA-BLG. Away from the half-filling, we have always the interlayer antiferromagnetic order parameter with the complicated nature of the coupling potential. Moreover, this is out of the scope of the present paper, and we consider here only the intralayer antiferromagnetism.                                           
%
\subsection{\label{sec:Section_3_4} Self-consistency equations}
%
Here, we write the complete set of $5$-dimensional system of equations for the principal physical parameters $\mu$, $\Delta_{\rm{AFM}}$ $\delta{\bar{n}}$, $\Delta_{\rm{exc} \uparrow}$ and $\Delta_{\rm{exc} \downarrow}$. Those parameters are given by the equations Eq.(\ref{Equation_22}) (for $\Delta_{\rm AFM}$), Eq.(\ref{Equation_30}) (for $\Delta_{\rm exc \uparrow}$ and $\Delta_{\rm exc \downarrow}$), Eq.(\ref{Equation_53}) (for the chemical potential $\mu$ and the average of density difference function $\delta{\bar{n}}$). In the ${\bf{k}}$-space representation this system of equations reads as
\begin{eqnarray}
	&&\frac{1}{\kappa}=-\frac{1}{N}\sum_{{\bf{k}}}\sum^{4}_{i=1}\alpha_{i{\bf{k}}}n_{F}(-\xi_{i{\bf{k}}}),
	\nonumber\\
	&&\Delta_{\rm{AFM}}=-\frac{U}{2N}\sum_{{\bf{k}}}\sum^{4}_{i=1}\left(\alpha'_{i{\bf{k}}}-\alpha''_{i{\bf{k}}}\right)n_{F}(-\xi_{i{\bf{k}}}),
	\nonumber\\
	&&\delta{\bar{n}}=-\frac{4}{{N}}\sum_{{\bf{k}}}\sum^{4}_{i=1}\beta_{i{\bf{k}}}n_{F}(-\xi_{i{\bf{k}}}),
	\nonumber\\
	&&\Delta_{\rm {exc}\uparrow}=-\frac{W}{N}\sum_{{\bf{k}}}\sum^{4}_{i=1}\gamma_{i{\bf{k}}}n_{F}(-\xi_{i{\bf{k}}}),
	\nonumber\\
 &&\Delta_{\rm {exc}\downarrow}=-\frac{W}{N}\sum_{{\bf{k}}}\sum^{4}_{i=1}\gamma'_{i{\bf{k}}}n_{F}(-\xi_{i{\bf{k}}}).
	\label{Equation_58}
\end{eqnarray} 
Here, the function $n_{F}\left(x\right)$ is the Fermi-Dirac distribution function defined as $n_{F}\left(\varepsilon_{i}\right)=1/\left[1+e^{\left(\varepsilon_{i}-\mu\right)/T}\right]$. The arguments $\xi_{i{\bf{k}}}$ in the Fermi-Dirac distribution functions are given as $\xi_{i{\bf{k}}}=\mu-\varepsilon_{i{\bf{k}}}$.
The explicit form of the ${\bf{k}}$-dependent coefficients $\alpha_{i{\bf{k}}}$, $\alpha'_{i{\bf{k}}}$, $\alpha''_{i{\bf{k}}}$, $\beta_{i{\bf{k}}}$, $\gamma_{i{\bf{k}}}$ and $\gamma'_{i{\bf{k}}}$, entering in the right-hand side in Eq.(\ref{Equation_57}), are given in Appendix \ref{sec:Section_6}, at the end of the paper. 

After solving the system of equations, in Eq.(\ref{Equation_57}), we get valuable information about the complicated physical phenomena in the AA-BLG system such as the antiferromagnetism, collective excitations, excitonic pairing, etc. All these phenomena and associated physical energy scales will be discussed and analysed in this and next sections. We should mention here a very important physical consequence coming from the form of the quasiparticles energy spectra, obtained in Eq.(\ref{Equation_50}), in the Section \ref{sec:Section_3_3} related to the fact that the energy parameters $\varepsilon_{i\bf{k}}$ (with $i=1,...,4$), are reducing to the usual single-layer graphene's energy dispersion relations, if the parameter $d$, in the subsquares, in the expression in Eq.(\ref{Equation_50}), is vanishing, i.e., $d\neq0$. Indeed, the parameter, $d$, in Eq.(\ref{Equation_50}), is the subject of the usual quadratic equation with respect to the antiferromagnetic order parameter $\Delta_{\rm AFM}$. It can be rewritten in the form 
\begin{eqnarray}
d=A(\Delta_{\rm AFM}-\Delta_{01})(\Delta_{\rm AFM}-\Delta_{02}).
\label{Equation_59}
\end{eqnarray}
In fact, the equation $d=0$ admits two different solutions at the given two different values of the parameter $\Delta_{\rm AFM}$: $\Delta_{\rm AFM}=\Delta_{\rm 01}$ or $\Delta_{\rm AFM}=\Delta_{\rm 02}$. The physical consequence following from this is that the monolayer graphene's spectrum appears, from the excitation spectrum of the bilayer graphene. This means that we have two possible antiferromagnetic orderings at which the system AA-BLG behaves like the usual monolayer graphene. This is very interesting from the point of view of technical applications of the AA-BLG structure because its functionality, in the mode of the graphene, is very promising for transport measurements and quantum information. At those values of the antiferromagnetic order parameter, the electrical conductivity in the AA-BLG system could be considerably increased, approaching the electrical conductivity in the pure monolayer graphene system. The system of the self-consistent equations in Eq.(\ref{Equation_57}) could be solved numerically by applying the Newton's Hybrid algorithm which replaces calls to the Jacobian function by its finite difference approximation, based on the original Fortran's library MINPACK. 
%
\section{\label{sec:Section_4} Results and discussions}
\begin{figure}
	\begin{center}
		\includegraphics[scale=0.33]{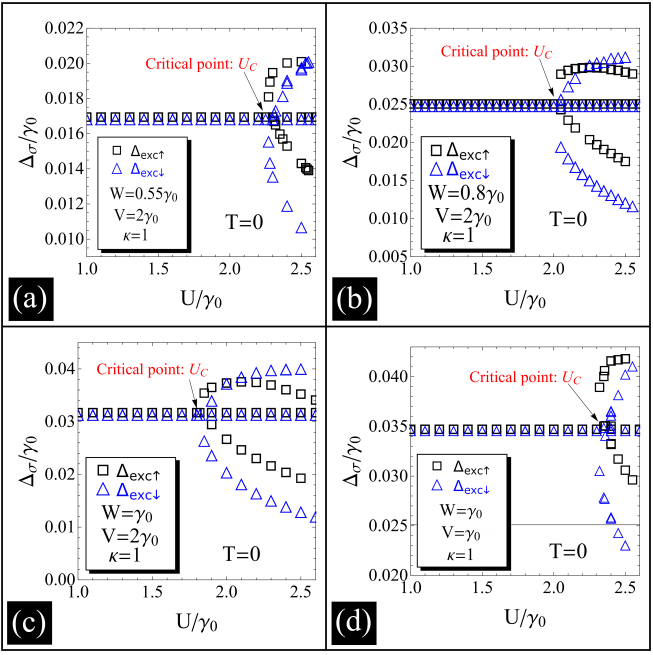}
		\caption{\label{fig:Fig_3}(Color online) The excitonic order parameters $\Delta_{{\rm exc}\uparrow}$ and $\Delta_{{\rm exc}\downarrow}$, as a function of the normalized intralayer Coulomb interaction parameter $U/\gamma_0$, for different values of the interlayer potential $W$ and external gate voltage $V$. The critical points $U=U_{C}$ are especially indicated in the figure which separate the single and triple valued regions of excitonic order parameters. The filling factor is set at the value $\kappa=1$ and the calculations are made for $T=0$.
}
\end{center}
\end{figure}

%
We present here the numerical results obtained by solving the system of equations, given in Eq.(\ref{Equation_57}). We calculate the principal physical parameters in the AA-BLG system such as the chemical potential $\mu$, the antiferromagnetic order parameter $\Delta_{\rm AFM}$, the average density difference function $\delta\bar{n}$ and the excitonic gap parameters $\Delta_{\rm exc \uparrow}$ and $\Delta_{\rm exc \downarrow}$. We consider their dependence on the Coulomb interaction parameter $U$ and the external electric field potential $V$. Four different values of $V$ have been considered in numerical calculations. All calculations have been performed for different reasonable values of the interlayer Coulomb interaction parameter $W$. In general, we estimate that
\begin{eqnarray}
W=(a_{0}/c_{0})U, 
\label{Equation_60}
\end{eqnarray}
where $c_{0}$ is the interlayer distance in the AA stacked BLG. The numerical results for the principal physical parameters $\mu$, $\delta{\bar{n}}$, $\Delta_{{\rm AFM}}$, $\Delta_{{\rm exc}\uparrow}$, and $\Delta_{{\rm exc}\downarrow}$ are given in Figs.~\ref{fig:Fig_2}-\ref{fig:Fig_7}. 
%
\subsection{\label{sec:Section_4_1} The staggered antiferromagnetic order}
\begin{figure}
	\begin{center}
		\includegraphics[scale=0.55]{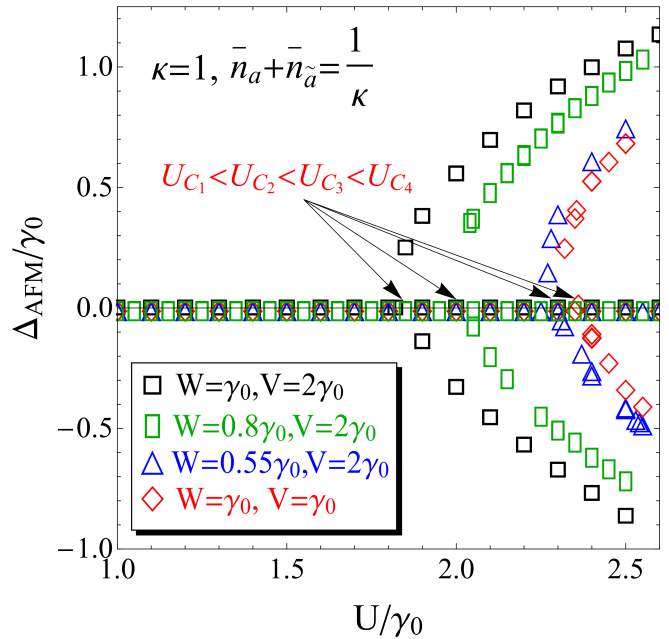}
		\caption{\label{fig:Fig_2}(Color online) The antiferromagnetic order parameter $\Delta_{\rm AFM}$, as a function of the normalized intralayer Coulomb interaction parameter $U/\gamma_0$. The curves in the picture (from left to right) were obtained for different values of the interlayer Coulomb interaction parameter $W$ and external electric field potential $V$, shown in the legend of the figure. The values of the critical points $U_{C}$ are shown in the figure by the arrows. The calculations are made for $T=0$. The inverse filling factor is set at the value $\kappa=1$ which corresponds to the partial filling in the layers.}
	\end{center}
\end{figure}

%
The solutions for the antiferromagnetic order parameter $\Delta_{\rm AFM}$ are presented in Fig.~\ref{fig:Fig_2}. The case of the partial filling is considered in Fig.~\ref{fig:Fig_2} with the inverse filling coefficient $\kappa=1$ (remember that the value $\kappa=0.5$ corresponds to the case of the half-filling in the layers of the AA-BLG. The $U$-dependence is shown in Fig.~\ref{fig:Fig_2}, for different values of the parameter $W$. We see that for all estimated values on the interlayer Coulomb interaction (see the values of the parameter $W$, given in the legend, in Fig.~\ref{fig:Fig_2}), the antiferromagnetic order parameter $\Delta_{\rm AFM}$ has a staggered behavior starting from some given point $U_{C}$, on the $U$-axis, which we call as the "critical value" of the on-site intralayer interaction parameter $U$. Different combinations of the interlayer Coulomb potential $W$ and applied gate voltage $V$ have been considered in  numerical evaluations, presented in Fig.~\ref{fig:Fig_2}, and the temperature is set at $T=0$. The first three curves, from left to right, were calculated for the fixed value of the external gate potential, set at $V=2\gamma_0=6$ eV and for different interlayer potential strengths: starting from the strong coupling at $W=\gamma_0$ to the weak coupling regime between the layers with $W=0.55\gamma_0$ (see the plot points in black, green and blue, in Fig.~\ref{fig:Fig_2}). The right outermost plot in red corresponds to the strong interlayer coupling (with $W=\gamma_0=3$ eV) and lower value of the gate potential $V$ ($V=\gamma_0=3$ eV). We observe, in Fig.~\ref{fig:Fig_2}, that all plots, obtained for the antiferromagnetic order parameter $\Delta_{\rm AFM}$ have a very important common feature that is related to the passage from the singlet to the triplet solutions at the certain critical value of the intralayer Coulomb potential $U=U_{C}$. For the values $U<U_{C}$ the antiferromagnetic order parameter vanishes and for $U>U_{C}$ (the above critical region) there are three distinct solutions for $\Delta_{\rm AFM}$: the positive branch with $\Delta_{\rm AFM}=\Delta^{+}_{\rm AFM}$, vanishing solution $\Delta_{\rm AFM}=0$, and the negative branch $\Delta_{\rm AFM}=\Delta^{-}_{\rm AFM}$. We see that the solution $\Delta_{\rm AFM}=0$ persists in the region $U>U_{C}$. The difference between positive and negative solution branches $\delta=\Delta^{+}_{\rm AFM}-\Delta^{-}_{\rm AFM}$ is increasing when augmenting the on-site Coulomb repulsion in the layers. We show here that such a triplet solution for the parameter $\Delta_{\rm AFM}$ is a direct consequence of the partial filling ($\kappa=1$). The maximum value for the absolute difference $\delta$ is attained in the region of the high values of the parameter $U$ and $\delta$ is smaller for the lower values of the interlayer potential $W$. The energy scale, related to $\delta$ (in the region of large-$U$), is situated in the range $\delta\in(3,6)$ eV. The smallest value for $\delta$ (with $\delta=1.165\gamma_0=3.497$ eV at $U=2.5\gamma_0$) is obtained for $W=\gamma_0=3$ eV and $V=\gamma_0=3$ eV, and the largest $\delta$ (with $\delta=1.938\gamma_{0}=5.814$) is obtained for $W=\gamma_0=3$ eV and $V=2\gamma_0=6$ eV. Taking into account the estimated values for $W$ and obtained values of $U_{C}$ we observe that the approximative estimation in Eq.(\ref{Equation_59}) works well with the replacement $U\rightarrow U_{C}$, for the values of $W$ and $V$, considered in panels (b) and (d), i.e., $W=(a_{0}/c_{0})U_{C}$. We see also a blue-shift effect for the critical value $U_{C}$ when decreasing the interlayer interaction parameter $W$ and applied gate potential $V$: $U_{C_{1}}<U_{C_{2}}<U_{C_{3}}<U_{C_{4}}$. This effect of the splitting of the antiferromagnetic order parameter is very analogue to the usual Stark-Lo Surdo effect caused by the presence of the applied electric field $V$. It is worth mentioning that the appearance of the antiferromagnetic ordering in the AA-BLG system opens the possibility to consider the magnetism in these structures. 

We will see furthermore, in this paper, that the behavior of the parameter $\Delta_{\rm AFM}$ is strongly governed by the average charge density imbalance $\delta{\bar{n}}=2\left(\bar{n}_{\tilde{a}}-\bar{n}_{a}\right)$ and also the excitonic order parameter $\Delta_{\rm exc}$. We suggest that all obtained values for $U_{C}$: $U_{C}\in\left(1.85\gamma_0,2.35\gamma_0\right)=\left(5.55,7.05\right)$ eV are situated in the energy interval which is in a good agreement with the estimated values of the on-site interaction $U$, given in Refs.\cite{cite_2, cite_45} and \cite{cite_46} could be attenuated experimentally by properly alternating the applied gate voltage and, thus, by fixing the interlayer potentials to the predicted values, given in Fig.~\ref{fig:Fig_2}. We will see, later on, that the same behavior of the parameter $\Delta_{\rm AFM}$ takes the place, if we consider dynamically changing $W$ with the approximative formula for the interlayer potential $W$: $W\approx \left(a_{0}/c_{0}\right)U$, where $a_{0}$ is the equilibrium carbon-carbon separation in the graphene's sheets ($a_{0}=1.42 \AA$) and $c_{0}$ is the interlayer separation ($c_{0}=3.35 \AA$). The transition obtained in Fig.~\ref{fig:Fig_2} is indeed a zero temperature transition from the non-magnetic phase (for the values $U<U_{C}$) to the antiferromagnetic one (for the values $U>U_{C}$) and is very similar with the Monte Carlo (\cite{cite_50}) and Hartree-Fock (\cite{cite_47, cite_48, cite_49}) theory results concerning the studies of the Hubbard model in the single honeycomb lattice of graphene. In those studies the transition to the antiferromagnetic insulator state in the single layer of graphene is obtained at some critical value of the on-site Coulomb repulsion much higher than the values of $U_{C}$. It will be clear hereafter that the antiferromagnetic states obtained here could be regarded as the excitonic antiferromagnetic insulator states because the triplet excitonic insulator phase appears at the same values of $U_{C}$ and coexists with the antiferromagnetic phase. This finding is very close to the results, given in Ref.\cite{cite_45}, where the antiferromagnetic phase is opening upon doping the system and the homogeneous metallic phase in the AA-BLG system becomes unstable with respect to the doping in the system. At the very high values of $U$ ($U>>U_{C}$), (we have not shown this result in Fig.~\ref{fig:Fig_2}), the staggered nature of the antiferromagnetic order parameter disappears and the metallic phase reentered, which is consistent with the results at large-$U$ in the Ref.\cite{cite_45}.  

On the other hand, the existence of the nono-zero values of the parameter $\Delta_{\rm AFM}$ means that above $U_{C}$ we have three channels for the polarization $\hat{p}_{z}$ of the electron gaz, i.e., the positive polarization $\hat{p}_{z}>0$, 0-polarization $\hat{p}_{z}=0$ and the negative polarization $\hat{p}_{z}<0$. The existence of non-zero polarization states in the AA-BLG system could lead to the spin-polarized currents in the system which is very promising for the applications of the considered AA-BLG system in the spintronics \cite{cite_17, cite_43}.
  
%
%
%
\subsection{\label{sec:Section_4_2} The excitonic order parameters}
%
The plots for the excitonic gap functions $\Delta_{\sigma}$ (for both spin directions $\sigma=\uparrow$ and $\sigma=\downarrow$) are presented in Fig.~\ref{fig:Fig_3} (see the panels (a), (b), (c) and (d)). The excitonic order parameters have been calculated self-consistently for different values of the repulsive interlayer interaction potential $W$, from weak to strong interaction limits. We observe in the panels (a)-(c) that there is a critical value of the intralayer Coulomb interaction parameter $U=U_{C}$ which separates the singlet and triplet solution regions and it is the case for both spin directions $\sigma=\uparrow$ and $\sigma=\downarrow$. Those critical values are exactly the same as in the case of the antiferromagnetic order parameter (see in Fig.~\ref{fig:Fig_2}). In the region $U<U_{C}$ we have $\Delta_{\uparrow}=\Delta_{\downarrow}$, while for $U>U_{C}$ there is a significant difference between them: $\delta_{\rm split}=\Delta_{\uparrow}-\Delta_{\downarrow}$. We see also that there is a red-shift effect, for the values $U_{C}$, in this case (when augmenting the parameter $W$ at the fixed value of the external electric field potential $V=2\gamma_0=6$ eV). This is opposite to the behavior of $U_{C}$, given in Fig.~\ref{fig:Fig_2}. Indeed, when augmenting the potential $W$, the point $U_{C}$ is displacing to the left on the $U$-axis (see in panels from (a) to (c), in Fig.~\ref{fig:Fig_3}). A small blue-shift effect is observed in the panel (d), in Fig.~\ref{fig:Fig_3}, for $W=\gamma_0=3$ eV and $V=\gamma_0=3$ eV. All points in the plots, in Fig.~\ref{fig:Fig_3}, have been calculated for the zero temperature limit, and the partial filling was considered there ($\kappa=1$).In difference with the antiferromagnetic order parameter $\Delta_{\rm AFM}$, the singlet excitonic order parameters, obtained in the region $U<U_{C}$, are not vanishing. The difference between the top and bottom solution branches for $\Delta_{\uparrow}$ and $\Delta_{\downarrow}$ shows the energy scale, related to the excitonic order in the system. After comparing the plots in Figs.~\ref{fig:Fig_2} and ~\ref{fig:Fig_3}, we see that the excitonic gaps $\Delta_{\uparrow}$ and $\Delta_{\downarrow}$ pass to the triplet solution region at the same value of the potential $U$ (i.e., $U_{C}$) at which the antiferromagnetic order parameter becomes non-zero. Therefore, we conclude that the antiferromagnetism in the AA-BLG system is strongly related to the excitonic pairing mechanism. It is worth to notice here that the coexisting region for the antiferromagnetic and excitonic phases is the region $U>U_{C}$ (above critical region). Moreover, we observe in Figs.~\ref{fig:Fig_2} and Fig.~\ref{fig:Fig_3} that the energy scale related to the antiferromagnetic order in the system is much larger than the energy scale related to the excitonic ordering, i.e., $|\Delta_{\rm AFM}|>>\Delta_{\sigma}$ (for both $\sigma=\uparrow$ and $\sigma=\downarrow$). The possible maximum value for the excitonic order parameter is obtained for $\sigma=\uparrow$ at $W=\gamma_0$ and $V=\gamma_0$ with $\Delta_{\uparrow\rm max}=0.042\gamma_0=126$ meV, while for the parameter $\Delta_{\rm AFM}$ we have $|\Delta_{\rm AFM}|_{\rm max}=0.6\gamma_0=1.8$ eV (at the same $W$ and $V$). The relative difference between them is of order of $\Delta_{\uparrow\rm max}/|\Delta_{\rm AFM}|_{\rm max}=7\times 10^{-2}$. Meanwhile, the absolute difference between the solutions $\Delta_{\uparrow}$ and $\Delta_{\downarrow}$ is relatively small. The maximum difference between $\Delta_{\uparrow}$ and $\Delta_{\downarrow}$ is attained at $W=\gamma_0=3$ eV and $V=\gamma_0=3$ eV. We have $|\Delta_{\uparrow}-\Delta_{\downarrow}|_{\rm max}=7.5\times 10^{-3}\gamma_0=2.5$ meV. In the region $U<U_{C}$, we get $|\Delta_{\sigma}|_{\rm max}=0.035\gamma_0=105$ meV, again at $W=\gamma_0$ and $V=\gamma_0$. 
%
\subsection{\label{sec:Section_4_3} The chemical potential and density imbalance}
\begin{figure}
	\begin{center}
		\includegraphics[scale=0.5]{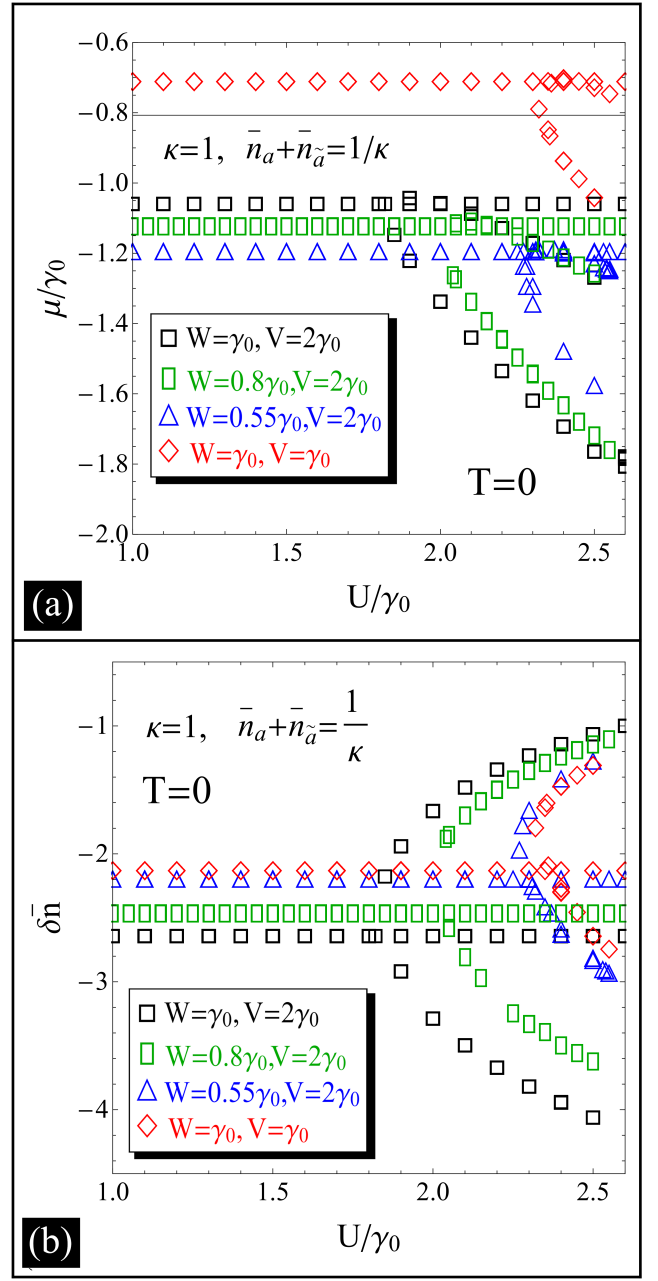}
		\caption{\label{fig:Fig_4}(Color online) Panel (a): The solution for the chemical potential (defined by the first equation in Eq.(\ref{Equation_57})), as a function of the normalized intralayer Coulomb interaction parameter $U/\gamma_0$, for different values of the interlayer potential $W$ and gate voltage $V$. Panel (b): The solution for the average charge density imbalance between the layers $\delta{\bar{n}}$, as a function of the normalized intralayer Coulomb interaction parameter $U/\gamma_0$. Different values of the interlayer potential $W$ and external gate potential $V$ are considered. The partial filling is considered here: $\kappa=1$ (this corresponds to the case of partial filling factor in the layers with $\kappa^{-1}=1$) and calculations are made for $T=0$.}
	\end{center}
\end{figure}

%
In the first panel (a), in Fig.~\ref{fig:Fig_4}, we have presented the numerical results for the chemical potential $\mu$ in the system (in our case of $T=0$, it gives also the Fermi level in the AA-BLG system for different values of the interaction parameter $U$) and in the second panel (b) we showed the solution for the average charge density imbalance $\delta{\bar{n}}$ between upper and lower layers in the AA-BLG construction.  
%
%
%
Different values of the parameter $W$ and external potential $V$ have been considered in panel (a), in Fig.~\ref{fig:Fig_4}. 
The general observation in Fig.~\ref{fig:Fig_4} is that the chemical potential $\mu$ is strictly negative ($\mu<0$) for all values of the on-site Coulomb repulsion $U$. 
This could be the sign for the excitonic condensate states in the AA-BLG under consideration. The chemical potential is multivalued (more precisely, triple-valued) above the critical points $U_{C}$, which are exactly the same as in the case of antiferromagnetic and excitonic order parameters (see in Figs.~\ref{fig:Fig_2} and  ~\ref{fig:Fig_3}). A  very interesting behavior is observed for the average charge density imbalance $\delta{\bar{n}}$ in the panel (b), in Fig.~\ref{fig:Fig_4}. Above the critical value $U_{C}$ it is triple valued and for the region $U<U_{C}$ it has only one single solution for each value of the on-site interaction parameter $U$. We will note the upper branch solution (for the values $U>U_{C}$) of $\delta{\bar{n}}$ as $\delta{\bar{n}}^{+}$, the lower branch solution as $\delta{\bar{n}}^{-}$ and the solution for $U<U_{C}$ as $\delta{\bar{n}}_{0}$. For the case $W=\gamma_0=3$ eV and $V=2\gamma_0=6$ eV (see the plot with the black square points, in Fig.~\ref{fig:Fig_2}) we get for $\delta{\bar{n}}_{0}$ the value $\delta{\bar{n}}_{0}=-2.647$, and we can calculate the corresponding average charge densities $\bar{n}_{a_{1}}$ and $\bar{n}_{{a}_{2}}$ after solving the system of algebraic equations in Eq.(\ref{Equation_53}). Then, we obtain $\bar{n}_{0a_{1}}=0.830875$ and $\bar{n}_{{0a}_{2}}=0.169125$. Thus, in the region $U<U_{C}$ the average population at the sites $A_{1}$, in the lower layer $\ell=1$, is much higher than the average population at the sites $A_{2}$, in the upper layer $\ell=2$. At the critical point $U=U_{C}=1.85\gamma_0=5.55$ eV, we get $\bar{n}_{Ca_{1}}=1$ and $\bar{n}_{Ca_{2}}=0$. This corresponds to the complete population inversion situation when the lower layer's sites $A_{1}$ are fully occupied by one electron (remember that we consider here the partial filling case with $\kappa=1$ at which the sum of average number of particles at the sites $A_{1}$ and $A_{2}$ is one). Such a population inversion corresponds to the electron-hole type and half-filled AA-BLG structure in which each lattice site $A_{1}$ and $B_{1}$ is occupied only be one electron and the upper layer sites are occupied by the holes. Then passing through the critical point $U_{C}$, towards the large values of $U$, we get, for example at $U=2.2\gamma_0=6.6$ eV: $\bar{n}^{+}_{a_{1}}=0.875$ and $\bar{n}^{+}_{a_{2}}=0.125$ and at $U=2.5\gamma_0=7.5$ eV: $\bar{n}^{+}_{a_{1}}=0.75$ and $\bar{n}^{+}_{a_{2}}=0.25$. We observe that the average number of population at the sites $A_{1}$ is diminishing and the average number of electron population at the sites $A_{2}$ is increasing, thus, the layer $\ell=2$ becomes populated with the electrons, more and more. Indeed, by choosing the appropriate values of the intralayer Coulomb interaction we can change the average number of electrons at the lattice sites in both layers and achieve the situation when the total population inversion takes place in the layers with $\bar{n}_{a_{1}}=1$ and $\bar{n}_{a_{2}}=0$. The blue-shift effect for the points $U_{C}$ also has the place for $\mu$ and $\delta{\bar{n}}$ when decreasing the interlayer Coulomb interaction parameter.

At the end of this subsection, we should emphasize the form of the chemical potential obtained here and especially its form in the upper critical region $U>U_{C}$. The chemical potential, being the energy scale for the creation or annihilation of a single particle in the system, is multivalued in the region $U>U_{C}$, shown in the upper panel (a), in Fig.~\ref{fig:Fig_4}), which means the existence of many possibilities or energetic excitation channels, for the creation of the excitonic pairs in the system. Note, that a single value of the chemical potential is responsible for the creation or annihilation of the excitonic pair. In turn, we suppose that structures of the antiferromagnetic order parameter $\Delta_{\rm AFM}$ (see in Fig.~\ref{fig:Fig_2}), the excitonic gap functions $\Delta_{\uparrow}$, $\Delta_{\downarrow}$ (see in Fig.~\ref{fig:Fig_3}) and the average charge density imbalance function $\delta{\bar{n}}$ (see in panel (b), in Fig.~\ref{fig:Fig_4}) are strongly correlated with the chemical potential behavior, and the later one governs their structure.

%
\subsection{\label{sec:Section_4_4} Interlayer potential}
\begin{figure}
	\begin{center}
		\includegraphics[scale=0.33]{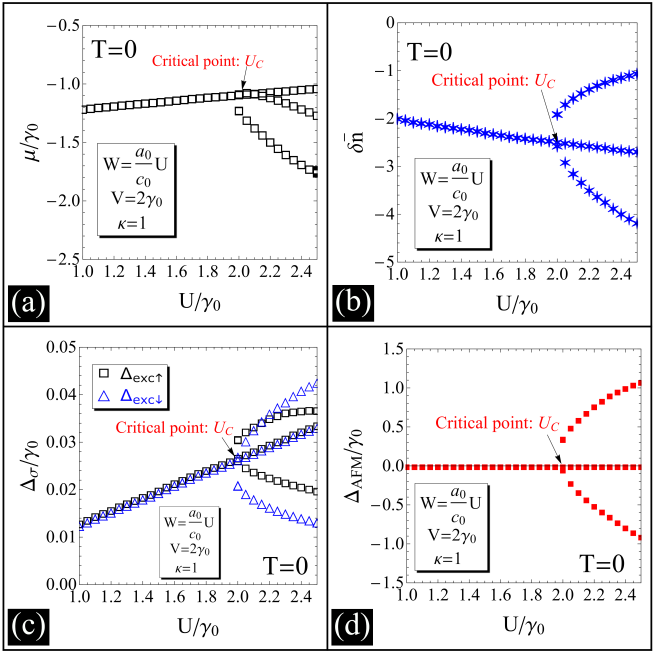}
		\caption{\label{fig:Fig_5}(Color online) Panel (a): The chemical potential in the AA-BLG system. Panel (b): The average charge density difference function between the layers. Panel (c): The excitonic order parameters $\Delta_{\uparrow}$ and $\Delta_{\downarrow}$. Panel (d): The antiferromagnetic order parameter $\Delta_{\rm AFM}$. All parameters have been obtained for the dynamical interlayer potential $W$ which varies from point to point with the intralayer Coulomb potential $U$: $W=\left(a_{0}/c_{0}\right)U$, where $c_{0}$ is the distance between the layers $c_{0}=3.35 \AA$. The parameter $\kappa$ is set at the value $\kappa=1$. The calculations are made for $T=0$.}
	\end{center}
\end{figure}

%
In Fig.~\ref{fig:Fig_5}, we have presented the solutions of the self-consistent equations in Eq.(\ref{Equation_58}) for the dynamically varying interlayer Coulomb interaction potential $W$ with the approximative formula $W=(a_{0}/c_{0})U$, discussed at the beginning in the Section \ref{sec:Section_4}. All calculations in Fig.~\ref{fig:Fig_5} have been done for the fixed value of the external electric gate potential $V=2\gamma_0=6$ eV and at $\kappa=1$. For the parameters $\mu$, $\delta{\bar{n}}$ and $\Delta_{\sigma}$, we have obtained the same general behavior as it was discussed in Sections \ref{sec:Section_4_1}, \ref{sec:Section_4_2} and \ref{sec:Section_4_3}. The difference between them lays in the solutions for the region $U<U_{C}$. In the present case the solution for the region $U<U_{C}$ vary when increasing the on-site potential $U$. Particularly, the chemical potential and the excitonic gap parameters are increasing with $U$ (see the plots in panels (a) and (c)), in the region $U<U_{C}$, while the average charge density imbalance $\Delta_{\sigma}$ is considerably decreasing (see in panel (b), in Fig.~\ref{fig:Fig_5}). This is not the case for the antiferromagnetic order parameter $\Delta_{\rm AFM}$, for which the zero line solution remains unchanged over the entire region of variation of the potential $U$. The dynamical passage from the singlet to the triplet solution states also remains unchanged for all parameters, presented in Fig.~\ref{fig:Fig_5}. The critical value, found for the on-site Coulomb potential $U$ is of order of $U_{C}=2.1\gamma_0=6.3$ eV (see the black arrows which mark the critical points, in Fig.~\ref{fig:Fig_5}).   
%
%
%
\subsection{\label{sec:Section_4_5} Dependence on the external electric field potential}
\begin{figure}
	\begin{center}
		\includegraphics[scale=0.33]{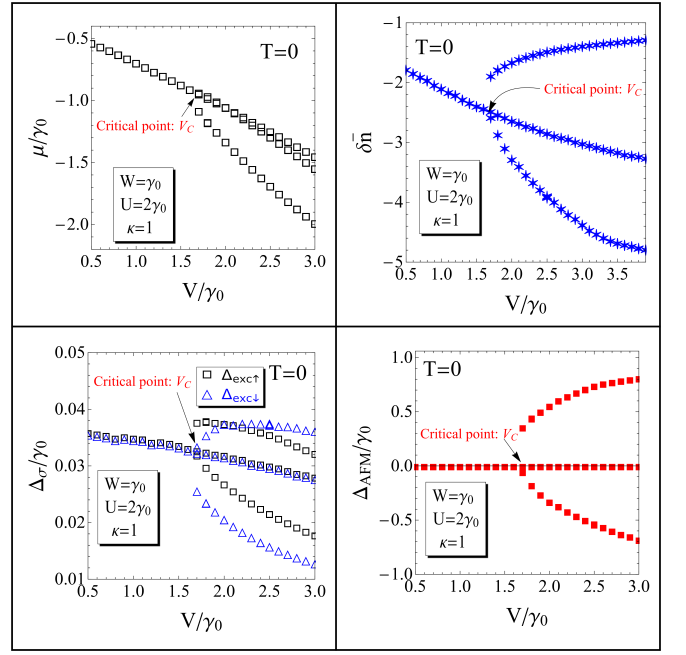}
		\caption{\label{fig:Fig_6}(Color online) The dependence of the physical parameters in the system on the applied external electric field potential $V$, for $W=\gamma_0=3$ eV and $U=2\gamma_0=6$ eV. The partial filling is considered here $\kappa=1$ and the temperature is set at $T=0$.}
	\end{center}
\end{figure}

\begin{figure}
	\begin{center}
		\includegraphics[scale=0.33]{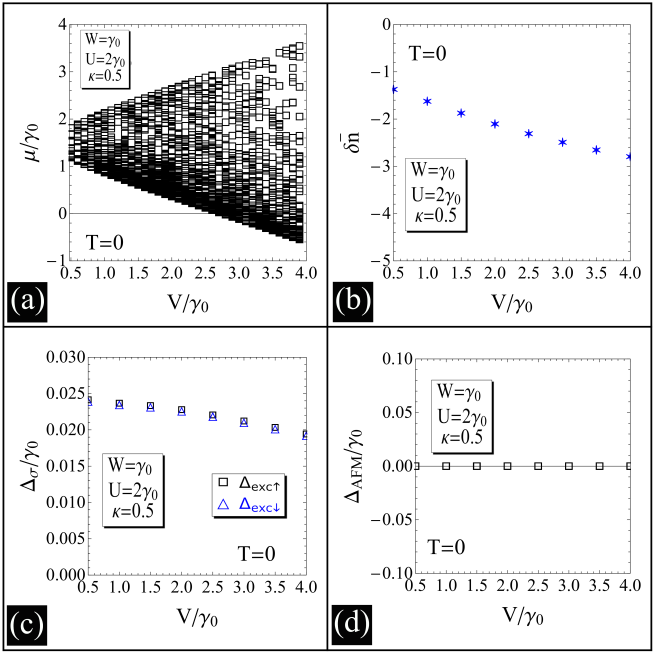}
		\caption{\label{fig:Fig_7}(Color online) The dependence of the physical parameters in the system on the applied external electric field potential $V$, for $W=\gamma_0=3$ eV and $U=2\gamma_0=6$ eV. The half-filling is considered here with $\kappa=0.5$, and the temperature is set at $T=0$.}
	\end{center}
\end{figure}

%
In Fig.~\ref{fig:Fig_6}, we have presented the $V$-dependence of all physical parameters in the system, calculated for $W=\gamma_0=3$ eV, $U=2\gamma_0=6$ eV and $\kappa=1$. A sufficiently large variation interval of $V$ has been considered $V\in(0.5\gamma_0,3\gamma_0)=(1.5,9)$ eV. We have obtained two critical regions for all parameters, as a function of the external gate potential. One, is the region before the critical point $V_{C}$ ($V<V_{C}$) and in this region all parameters decrease when increasing the potential $V$ (see the plots (a), (b) and (c), in Fig.~\ref{fig:Fig_6}), except the antiferromagnetic order parameter $\Delta_{\rm AFM}$ (see in plot (d), in Fig.~\ref{fig:Fig_6}) with the solution line $\Delta_{\rm AFM}=0$ which remains unchanged until the value $U=U_{C}$. The other region is the above critical region $V>V_{C}$ where all parameters are triple valued and the transition from the singlet to the triplet regions occurs at $V_{C}=1.7\gamma_0=5.1$ eV.  
%
%
%
Another interesting behavior of the physical parameters as a function of applied gate potential $V$ was observed for the case of the half-filling, i.e. $\kappa=0.5$ (this is shown in Fig.~\ref{fig:Fig_7}). The parameters $W$ and $U$ were fixed to the same values as in Fig.~\ref{fig:Fig_6} and the same interval of variation of the external gate potential $V$ was considered $V\in(0.5\gamma_0,3\gamma_0)=(1.5,4.5)$ eV. In this case, all parameters, except the chemical potential are single-valued over the whole region of variation of the potential $V$ (see in panels (b), (c) and (d)). The charge density function $\delta{\bar{n}}$ and excitonic gap parameters decrease when increasing the potential $V$ (see in panels (b) and (c), in Fig.~\ref{fig:Fig_6}). Moreover, the excitonic gaps for different spin orientations are coinciding in this case, i.e., $\Delta_{\uparrow}=\Delta_{\downarrow}$. The chemical potential shows a large band structure (see in panel (a), in Fig.~\ref{fig:Fig_7}) in the mentioned region of variation of the external electric field and there is no critical point of transition, in this case. The antiferromagnetic order parameter is zero, for all values of $V$. We suppose that the result $\Delta_{\rm AFM}=0$ is a direct consequence of the absence of the critical point of transition on the $V$-axis. In our strong conviction, the behavior of the parameters $\Delta_{\sigma}$ and $\Delta_{\rm AFM}$ is governed by the function $\delta{\bar{n}}$. Moreover, we are convinced that the appearance of the antiferromagnetic order in the system is strongly related to the existence of the critical point of transition from the single-valued to the triple valued states. Nevertheless, the excitonic ordering exists in Fig.~\ref{fig:Fig_7} and it could be more stable because of the behavior of the chemical potential, shown in panel (a), in Fig.~\ref{fig:Fig_7}.
%
%
%
\section{\label{sec:Section_5} Concluding remarks}
%
We have studied the antiferromagnetism and excitonic pair formations in the AA-stacked bilayer graphene. The on-site intralayer and local interlayer Coulomb interactions have been considered within the bilayer Hubbard model. We have shown that the antiferromagnetic and excitonic phases can coexist only in the regime away from the half-filling case, and after considering the partial filling in the layers. We obtained a system of coupled self-consistent equations and solved it exactly by applying a finite difference version of the Hybrid algorithm without internal scaling. We calculated numerically the chemical potential, excitonic order parameters, average particle population difference between the layers and the antiferromagnetic order parameter. We obtained the dependence of those physical parameters on the on-site Coulomb repulsion $U$, for the fixed and dynamic values of the interlayer potential $W$, in the presence of the external gate potential $V$. Furthermore, we considered also the $V$-dependence of those calculated quantities. We found the critical values of the potentials $U$ and $V$, which separate two types of solution regions: single-valued and triple-valued, for all calculated parameters. Especially, we have shown that the antiferromagnetic order parameter is zero in the regions $U<U_{C}$ and $V<V_{C}$, and above those values, i.e., when $U>U_{C}$ and $V>V_{C}$, a Stark-Lo Surdo type of splitting effect takes place for $\Delta_{\rm AFM}$. Therefore, the antiferromagnetism is completely absent below the critical values $U_{C}$ and $V_{C}$. The same type of effect was shown also for the other calculated parameters in the system with a difference of the non-zero solutions in the regions $U<U_{C}$ and $V<V_{C}$.
Next, we considered also the case of the half-filling regime, and we calculated the $V$ dependence in this case. As the numerical calculations show, all triplet solutions appear in the case away from the half-filling and can be tuned by the external electric field potential $V$. Especially, the existence of the critical point, which separates the single-valued and triple-valued solutions is a direct consequence of the partial filling in the layers and the antiferromagnetic staggered order appear only in the case of the partial filling. From the results in the paper, it is clear that the behavior of the excitonic and antiferromagnetic order parameters is strongly governed by the average particle population difference function between the layers. The most important result in the present paper is related to the coexistence of the antiferromagnetic and excitonic phases in the system. We have shown that they can coexist only in the region above critical values $U_{C}$ and $V_{C}$. In the case of half-filling, the excitonic phase exists without the antiferromagnetic one and this is the consequence of the absence of the critical point of transition from the singlet solutions to the triplet ones. The obtained antiferromagnetic phase could be considered as the excitonic antiferromagnetic insulator state. On the other hand, the existence of non-zero antiferroamgnetic (spin-polarization) states in the AA-BLG system could lead to the spin-polarized currents in the system, which is very promising for the applications of the considered AA-BLG system in spintronics. We have calculated the important energy scales in the AA-BLG system, by showing that the antiferromagnetic order parameter is much larger than the excitonic one and we have estimated the order of magnitudes of those order parameters. Moreover, we have shown that at the critical values $U_{C}$ and $V_{C}$ and at partial filling, the bilayer graphene shows the properties of the electron-hole type of bilayer at the half-filling, where there is only one electron at the given lattice site in the lower layer and one hole at the adjacent lattice site position in the upper layer. When passing through the critical points towards the triplet solution region this behavior was changing, i.e., the upper layer becomes more and more populated and the lower layer is emptying. Thus, the obtained critical points could be considered as the points at which the average particle population inversion occurs in the bilayer graphene. This process could be controlled by the applied external gate potential. All considered values of external gate potential and interaction parameters could be experimentally accessible, and the interlayer interaction potential could be fixed or tuned also via the applied gate voltage.   

The results obtained in this paper could be important in the situations when the AA-BLG construction is investigated experimentally for the spin-controlled electronic transport, and also for the multichannel coherent excitonic states in the double bilayer heterojunctions \cite{cite_20}. The important energy scales and their inter-relations, obtained here for AA-BLG construction, will enhance and simplify further investigations, both theoretical and experimental, on the structures, based on AA-BLG systems. As the next step of our calculations, we plan to consider the influence of the antiferromagnetic and excitonic states on the electronic band structure in the AA-BLG system, under the present study. The other two important directions of evaluations of the presented here theory are related to the problems of detection of the total spin polarization in the layers in AA-BLG and also the consideration of the $(p-n)_{2}-(n-p)_{1}$ type square heterojunctions (where the index $2$ indicates the layer $2$ and the index $1$ the layer $1$), based on the double-bilayer graphene, where the stable intralayer and interlayer excitonic states could be created. These last heterostructures could be very important constructions for the creation of a new generation of amplifiers for the storage of solar energy, thermal rectifiers and batteries and could lead to a new type of optoelectronics and photonics.       

The examination of the stable antiferromagnetic ordering in AA-BLG could have a tremendous significance for realizing the room-temperature spintronic memories and nanoscale tunnel junctions\cite{cite_56, cite_57} with the negligible stray fields \cite{cite_58}.
\appendix
%
\section{\label{sec:Section_6} The coefficients in the polynomials}
%
Here, we present the analytical forms of the ${\bf{k}}$-dependent coefficients $\alpha_{i{\bf{k}}}$, $\alpha'_{i{\bf{k}}}$, $\alpha''_{i{\bf{k}}}$, $\beta_{i{\bf{k}}}$, $\gamma_{i{\bf{k}}}$ and $\gamma'_{i{\bf{k}}}$, in Eq.(\ref{Equation_57}) which are represented in terms of the polynomial functions in terms of the quasiparticle excitation energies $\varepsilon_{i{\bf{k}}}$ (with $i=1,...4$), in Eq.(\ref{Equation_50}), in the Section \ref{sec:Section_3_3}. Namely, we have $\Delta_{\rm AFM}=0$, for $U<U_{C}$, $\Delta_{\rm AFM}\neq0$, for $U>U_{C}$ $\exists U_{C}$ critical

\begin{eqnarray}
\footnotesize
\arraycolsep=0pt
\medmuskip = 0mu
\alpha_{i{{\bf{k}}}}
=(-1)^{i+1}
\left\{
\begin{array}{cc}
\displaystyle  & \frac{{\cal{P}}^{(3)}_{1}(\varepsilon_{i{\bf{k}}})}{\left(\varepsilon_{1{\bf{k}}}-\varepsilon_{2{\bf{k}}}\right)}\prod^{}_{j=3,4}\frac{1}{\left(\varepsilon_{i{\bf{k}}}-\varepsilon_{j{\bf{k}}}\right)},  \ \ \  $if$ \ \ \ i=1,2,
\newline\\
\newline\\
\displaystyle  & \frac{{\cal{P}}^{(3)}_{1}(\varepsilon_{i{\bf{k}}})}{\left(\varepsilon_{3{\bf{k}}}-\varepsilon_{4{\bf{k}}}\right)}\prod^{}_{j=1,2}\frac{1}{\left(\varepsilon_{i{\bf{k}}}-\varepsilon_{j{\bf{k}}}\right)},  \ \ \  $if$ \ \ \ i=3,4.
\end{array}\right.
\nonumber\\
\label{Equation_A_1}
\end{eqnarray}
Here, the function ${\cal{P}}^{(3)}_{1}(x)$ is the polynomial of third order and is defined as
\begin{eqnarray}
{\cal{P}}^{(3)}_{1}(x)=A_{1}x^{3}+B_{1}x^{2}+C_{1}x+D_{1},
\label{Equation_A_2}
\end{eqnarray}  
where the coefficients $A_{1}, B_{1}, C_{1}$ and $D_{1}$ are given as
\begin{eqnarray}
&&A_{1}=4,
\nonumber\\
&&B_{1}=12W-6\left(\mu_{1}+\mu_{2}\right),
\nonumber\\
&&C_{1}=8W^{2}-4|\tilde{\gamma}_{{\bf{{k}}}}|^{2}-2\left(\gamma_{1}+\Delta_{\rm {exc}\uparrow}\right)^{2}
-2\left(\gamma_{1}+\Delta_{\rm {exc}\downarrow}\right)^{2}
\nonumber\\
&&-4\Delta^{2}_{{\rm AFM}}-16W\mu_{1}-8W\mu_{2},
\nonumber\\
&&D_{1}=\left(2W-\mu_{1}-\mu_{2}\right)\left[2\mu_{1}\mu_{2}-\left(\gamma_{1}+\Delta_{\rm{exc}\uparrow}\right)^{2}-2\Delta^{2}_{{\rm AFM}}\right.
\nonumber\\
&&\left.-2|\tilde{\gamma}_{0{\bf{k}}}|^{2}-4W\mu_{1}\right].
\label{Equation_A_3}
\end{eqnarray} 
The coefficients $\alpha'_{i{\bf{k}}}$ and $\alpha''_{i{\bf{k}}}$, in the second equation in Eq.(\ref{Equation_57}), are defined in the form
\begin{eqnarray}
\footnotesize
\arraycolsep=0pt
\medmuskip = 0mu
\alpha'_{i{{\bf{k}}}}
=(-1)^{i+1}
\left\{
\begin{array}{cc}
\displaystyle  & \frac{{\cal{P}}^{(3)}_{2}(\varepsilon_{i{\bf{k}}})}{\left(\varepsilon_{1{\bf{k}}}-\varepsilon_{2{\bf{k}}}\right)}\prod^{}_{j=3,4}\frac{1}{\left(\varepsilon_{i{\bf{k}}}-\varepsilon_{j{\bf{k}}}\right)},  \ \ \  $if$ \ \ \ i=1,2,
\newline\\
\newline\\
\displaystyle  & \frac{{\cal{P}}^{(3)}_{2}(\varepsilon_{i{\bf{k}}})}{\left(\varepsilon_{3{\bf{k}}}-\varepsilon_{4{\bf{k}}}\right)}\prod^{}_{j=1,2}\frac{1}{\left(\varepsilon_{i{\bf{k}}}-\varepsilon_{j{\bf{k}}}\right)},  \ \ \  $if$ \ \ \ i=3,4,
\end{array}\right.
\nonumber\\
\label{Equation_A_4}
\end{eqnarray}
and
\begin{eqnarray}
\footnotesize
\arraycolsep=0pt
\medmuskip = 0mu
\alpha''_{i{{\bf{k}}}}
=(-1)^{i+1}
\left\{
\begin{array}{cc}
\displaystyle  & \frac{{\cal{P}}^{(3)}_{3}(\varepsilon_{i{\bf{k}}})}{\left(\varepsilon_{1{\bf{k}}}-\varepsilon_{2{\bf{k}}}\right)}\prod^{}_{j=3,4}\frac{1}{\left(\varepsilon_{i{\bf{k}}}-\varepsilon_{j{\bf{k}}}\right)},  \ \ \  $if$ \ \ \ i=1,2,
\newline\\
\newline\\
\displaystyle  & \frac{{\cal{P}}^{(3)}_{3}(\varepsilon_{i{\bf{k}}})}{\left(\varepsilon_{3{\bf{k}}}-\varepsilon_{4{\bf{k}}}\right)}\prod^{}_{j=1,2}\frac{1}{\left(\varepsilon_{i{\bf{k}}}-\varepsilon_{j{\bf{k}}}\right)},  \ \ \  $if$ \ \ \ i=3,4.
\end{array}\right.
\nonumber\\
\label{Equation_A_5}
\end{eqnarray}
Here, the polynomials ${\cal{P}}^{(3)}_{2}(x)$ and ${\cal{P}}^{(3)}_{3}(x)$ are the polynomials of third order with respect to the argument $x$. They are given by the relations 
\begin{eqnarray}
{\cal{P}}^{(3)}_{2}(x)=A_{2}x^{3}+B_{2}x^{2}+C_{2}x+D_{2},
\nonumber\\
{\cal{P}}^{(3)}_{3}(x)=A_{3}x^{3}+B_{3}x^{2}+C_{3}x+D_{3},
\label{Equation_A_6}
\end{eqnarray}  
and the parameters $A_{i}, B_{i}, C_{i}$ and $D_{i}$ with $i=2,3$, in Eq.(\ref{Equation_A_6}), are defined by the relations 
\begin{eqnarray}
&&A_{2}=1,
\nonumber\\
&&B_{2}=4W+\Delta_{\rm{AFM}}-\mu_{1}-2\mu_{2},
\nonumber\\
&&C_{2}=4W^{2}-|\tilde{\gamma}_{{\bf{{k}}}}|^{2}-\left(\gamma_{1}+\Delta_{\rm {exc}\downarrow}\right)^{2}
+4W\Delta_{{\rm AFM}}-\Delta^{2}_{{\rm AFM}},
\nonumber\\
&&-4W\left[\mu_{1}+\mu_{2}\right]-2\Delta_{\rm{AFM}}\mu_{2}+2\mu_{1}\mu_{2}+\mu^{2}_{2},
\nonumber\\
&&D_{2}=-\left(2W+\Delta_{\rm{AFM}}-\mu_{2}\right)\left(2\gamma_{1}\Delta_{{\rm exc}\downarrow}+\gamma^{2}_{1}+\Delta^{2}_{{\rm exc}\downarrow}\right)
\nonumber\\
&&-\left(\Delta_{\rm{AFM}}-\mu_{1}\right)\left[|\tilde{\gamma}_{\bf{{k}}}|^{2}+\Delta^{2}_{\rm {AFM}}-\left(2W-\mu_{2}\right)^{2}\right],
\label{Equation_A_7}
\end{eqnarray} 
and
\begin{eqnarray}
&&A_{3}=1,
\nonumber\\
&&B_{3}=4W-\Delta_{{\rm AFM}}-\mu_{1}-2\mu_{2},
\nonumber\\
&&C_{3}=4W^{2}-|\tilde{\gamma}_{{\bf{{k}}}}|^{2}-2\left(\gamma_{1}+\Delta_{\rm {exc}\uparrow}\right)^{2}
-4W\Delta_{{\rm AFM}}-\Delta^{2}_{{\rm AFM}}
\nonumber\\
&&-4W\Delta_{{\rm AFM}}-4W\left(\mu_{1}+\mu_{2}\right)+2{\Delta_{{\rm AFM}}}\mu_{2}+2\mu_{1}\mu_{2}+\mu^{2}_{2},
\nonumber\\
&&D_{3}=-\left(2W-\Delta_{{\rm AFM}}-\mu_{2}\right)\left(2\gamma_1\Delta_{{\rm exc},\uparrow}+\gamma^{2}_{1}+\Delta^{2}_{{\rm exc}\uparrow}\right)
\nonumber\\
&&-\left(-\mu_{1}-\Delta_{{\rm AFM}}\right)\left(|\tilde{\gamma}_{{\bf{{k}}}}|^{2}+\Delta^{2}_{{\rm AFM}}-\left(2W-\mu_{2}\right)^{2}\right).
\label{Equation_A_8}
\end{eqnarray} 
As we see here, the coefficients $C_{3}$ and $D_{3}$ in Eq.(\ref{Equation_A_8}) could be obtained from the coefficients $C_{2}$ and $D_{2}$ in Eq.(\ref{Equation_A_7}) just by the replacing $\Delta_{{\rm exc}\uparrow}\rightleftharpoons\Delta_{{\rm exc}\downarrow}$ and $\Delta_{{\rm AFM}}\rightleftharpoons -\Delta_{{\rm AFM}}$. Furthermore, the coefficients $\beta_{i{\bf{k}}}$ entering in the right-hand side in the equation for $\delta{\bar{n}}$ (see, the third equation, in Eq.(\ref{Equation_57}), are
\begin{eqnarray}
\footnotesize
\arraycolsep=0pt
\medmuskip = 0mu
\beta_{i{{\bf{k}}}}
=(-1)^{i+1}
\left\{
\begin{array}{cc}
\displaystyle  & \frac{{\cal{P}}^{(2)}_{4}(\varepsilon_{i{\bf{k}}})}{\left(\varepsilon_{1{\bf{k}}}-\varepsilon_{2{\bf{k}}}\right)}\prod^{}_{j=3,4}\frac{1}{\left(\varepsilon_{i{\bf{k}}}-\varepsilon_{j{\bf{k}}}\right)},  \ \ \  $if$ \ \ \ i=1,2,
\newline\\
\newline\\
\displaystyle  & \frac{{\cal{P}}^{(2)}_{4}(\varepsilon_{i{\bf{k}}})}{\left(\varepsilon_{3{\bf{k}}}-\varepsilon_{4{\bf{k}}}\right)}\prod^{}_{j=1,2}\frac{1}{\left(\varepsilon_{i{\bf{k}}}-\varepsilon_{j{\bf{k}}}\right)},  \ \ \  $if$ \ \ \ i=3,4
\end{array}\right.
\nonumber\\
\label{Equation_A_9}
\end{eqnarray}
with the second order polynomial ${\cal{P}}^{(2)}_{4}(\varepsilon_{i{\bf{k}}})=A_{4}x^{2}+B_{4}x+C_{4}$, where the coefficients $A_{4}$, $B_{4}$ and $C_{4}$ have the following form 
\begin{eqnarray}
&&A_{4}=-2W-\mu_{1}+\mu_{2},
\nonumber\\
&&B_{4}=-4W^{2}+\left(\gamma_{1}+\Delta_{{\rm exc}\uparrow}\right)^{2}-\left(\gamma_{1}+\Delta_{{\rm exc}\downarrow}\right)^{2}+4W\Delta_{{\rm AFM}}
\nonumber\\
&&+2\Delta_{{\rm AFM}}\left(\mu_{1}-\mu_{2}\right)+4W\mu_{2}+\mu^{2}_{1}-\mu^{2}_{2},
\nonumber\\
&&C_{4}=\left(-2W-\mu_{1}+\mu_{2}\right)|\tilde{\gamma}_{{\bf{{k}}}}|^{2}+\left(\gamma_1+\Delta_{{\rm AFM}\uparrow}\right)^{2}\times
\nonumber\\
&&\left(2W-\Delta_{\rm AFM}-\mu_{2}\right)+\left(\gamma_1+\Delta_{{\rm exc}\uparrow}\right)^{2}\left(\Delta_{{\rm AFM}}+\mu_{1}\right)
\nonumber\\
&&+4W^{2}\left(\Delta_{{\rm AFM}}+\mu_{1}\right)-2W\left(\Delta^{2}_{\rm AFM}-\mu^{2}_{1}+2\Delta_{\rm AFM}\mu_{2}\right.
\nonumber\\
&&\left.+2\mu_{1}\mu_{2}\right)-\Delta^{2}_{\rm AFM}\left(\mu_{1}-\mu_{2}\right)-\Delta_{\rm AFM}\left(\mu^{2}_{1}-\mu^{2}_{2}\right)
\nonumber\\
&&-\mu_{1}\mu_{2}\left(\mu_{1}-\mu_{2}\right).
\label{Equation_A_10}
\end{eqnarray} 
Next, the coefficients $\gamma_{i{\bf{k}}}$ and $\gamma'_{i{\bf{k}}}$, in the last two equations, in Eq.(\ref{Equation_57}), are given with the help of the second order polynomials ${\cal{P}}^{(2)}_{5}(x)=A_{5}x^{2}+B_{5}x+C_{5}$ and ${\cal{P}}^{(2)}_{6}(x)=A_{6}x^{2}+B_{6}x+C_{6}$. We have 
\begin{eqnarray}
\footnotesize
\arraycolsep=0pt
\medmuskip = 0mu
\gamma_{i{{\bf{k}}}}
=(-1)^{i+1}
\left\{
\begin{array}{cc}
\displaystyle  & \frac{{\cal{P}}^{(2)}_{5}(\varepsilon_{i{\bf{k}}})}{\left(\varepsilon_{1{\bf{k}}}-\varepsilon_{2{\bf{k}}}\right)}\prod^{}_{j=3,4}\frac{1}{\left(\varepsilon_{i{\bf{k}}}-\varepsilon_{j{\bf{k}}}\right)},  \ \ \  $if$ \ \ \ i=1,2,
\newline\\
\newline\\
\displaystyle  & \frac{{\cal{P}}^{(2)}_{5}(\varepsilon_{i{\bf{k}}})}{\left(\varepsilon_{3{\bf{k}}}-\varepsilon_{4{\bf{k}}}\right)}\prod^{}_{j=1,2}\frac{1}{\left(\varepsilon_{i{\bf{k}}}-\varepsilon_{j{\bf{k}}}\right)},  \ \ \  $if$ \ \ \ i=3,4,
\end{array}\right.
\nonumber\\
\label{Equation_A_11}
\end{eqnarray}
and
\begin{eqnarray}
\footnotesize
\arraycolsep=0pt
\medmuskip = 0mu
\gamma'_{i{{\bf{k}}}}
=(-1)^{i+1}
\left\{
\begin{array}{cc}
\displaystyle  & \frac{{\cal{P}}^{(2)}_{6}(\varepsilon_{i{\bf{k}}})}{\left(\varepsilon_{1{\bf{k}}}-\varepsilon_{2{\bf{k}}}\right)}\prod^{}_{j=3,4}\frac{1}{\left(\varepsilon_{i{\bf{k}}}-\varepsilon_{j{\bf{k}}}\right)},  \ \ \  $if$ \ \ \ i=1,2,
\newline\\
\newline\\
\displaystyle  & \frac{{\cal{P}}^{(2)}_{6}(\varepsilon_{i{\bf{k}}})}{\left(\varepsilon_{3{\bf{k}}}-\varepsilon_{4{\bf{k}}}\right)}\prod^{}_{j=1,2}\frac{1}{\left(\varepsilon_{i{\bf{k}}}-\varepsilon_{j{\bf{k}}}\right)}.  \ \ \  $if$ \ \ \ i=3,4.
\end{array}\right.
\nonumber\\
\label{Equation_A_12}
\end{eqnarray}
The coefficients $A_{5}$, $B_{5}$, $C_{5}$, $A_{6}$, $B_{6}$ and $C_{6}$ are the subject of the following expressions
\begin{eqnarray}
&&A_{5}=\gamma_{1}+\Delta_{{\rm exc}\uparrow},
\nonumber\\
&&B_{5}=\left(\gamma_{1}+\Delta_{{\rm exc}\uparrow}\right)\left(2W-\mu_{1}-\mu_{2}\right),
\nonumber\\
&&C_{5}=\left(\gamma_1+\Delta_{{\rm exc}\downarrow}\right)\left[|\tilde{\gamma}_{{\bf{{k}}}}|^{2}-\left(\gamma_1+\Delta_{{\rm exc}\uparrow}\right)\left(\gamma_1+\Delta_{{\rm exc}\downarrow}\right)\right]
\nonumber\\
&&+\left(\gamma_1+\Delta_{{\rm exc}\uparrow}\right)\left(\Delta_{{\rm AFM}}-\mu_{1}\right)\left(2W-\Delta_{\rm AFM}-\mu_{2}\right)
\label{Equation_A_13}
\end{eqnarray} 
and
\begin{eqnarray}
&&A_{6}=\gamma_{1}+\Delta_{{\rm exc}\downarrow},
\nonumber\\
&&B_{6}=\left(\gamma_{1}+\Delta_{{\rm exc}\downarrow}\right)\left(2W-\mu_{1}-\mu_{2}\right),
\nonumber\\
&&C_{6}=\left(\gamma_1+\Delta_{{\rm exc}\uparrow}\right)\left[|\tilde{\gamma}_{{\bf{{k}}}}|^{2}-\left(\gamma_1+\Delta_{{\rm exc}\downarrow}\right)\left(\gamma_1+\Delta_{{\rm exc}\uparrow}\right)\right]
\nonumber\\
&&+\left(\gamma_1+\Delta_{{\rm exc}\downarrow}\right)\left(-\Delta_{{\rm AFM}}-\mu_{1}\right)\left(2W+\Delta_{\rm AFM}-\mu_{2}\right).
\nonumber\\
\label{Equation_A_14}
\end{eqnarray} 
It is clear from Eqs.(\ref{Equation_A_14}) and (\ref{Equation_A_13}) that the parameters $A_{6},B_{6},C_{6}$ could be obtained from the parameters $A_{5},B_{5},C_{5}$ just by the replacements $\Delta_{{\rm exc}\uparrow}\rightleftharpoons\Delta_{{\rm exc}\downarrow}$ and $\Delta_{{\rm AFM}}\rightleftharpoons-\Delta_{{\rm AFM}}$.

\section*{References}


\begin{thebibliography}{58}
	\providecommand{\natexlab}[1]{#1}
	\providecommand{\url}[1]{\texttt{#1}}
	\expandafter\ifx\csname urlstyle\endcsname\relax
	\providecommand{\doi}[1]{doi: #1}\else
	\providecommand{\doi}{doi: \begingroup \urlstyle{rm}\Url}\fi
	
	\bibitem[McCann and Koshino(2013)]{cite_1}
	Edward McCann and Mikito Koshino.
	\newblock The electronic properties of bilayer graphene.
	\newblock \emph{Reports on Progress in Physics}, 76\penalty0 (5):\penalty0
	056503, apr 2013.
	\newblock \doi{10.1088/0034-4885/76/5/056503}.
	
	\bibitem[Rozhkov et~al.(2016)Rozhkov, Sboychakov, Rakhmanov, and Nori]{cite_2}
	A.V. Rozhkov, A.O. Sboychakov, A.L. Rakhmanov, and Franco Nori.
	\newblock Electronic properties of graphene-based bilayer systems.
	\newblock \emph{Physics Reports}, 648:\penalty0 1--104, 2016.
	\newblock ISSN 0370-1573.
	\newblock \doi{10.1016/j.physrep.2016.07.003}.
	\newblock Electronic properties of graphene-based bilayer systems.
	
	\bibitem[Junaid and Witjaksono(2019)]{cite_3}
	Muhammad Junaid and Gunawan Witjaksono.
	\newblock Analysis of band gap in aa and ab stacked bilayer graphene by
	hamiltonian tight binding method.
	\newblock In \emph{2019 IEEE International Conference on Sensors and
		Nanotechnology}, pages 1--4, 2019.
	\newblock \doi{10.1109/SENSORSNANO44414.2019.8940102}.
	
	\bibitem[Zhang et~al.(2009{\natexlab{a}})Zhang, Tang, Girit, Hao, Martin,
	Zettl, Crommie, Shen, and Wang]{cite_4}
	Yuanbo Zhang, Tsung-Ta Tang, Caglar Girit, Zhao Hao, Michael~C Martin, Alex
	Zettl, Michael~F Crommie, Y~Ron Shen, and Feng Wang.
	\newblock Direct observation of a widely tunable bandgap in bilayer graphene.
	\newblock \emph{Nature}, 459\penalty0 (7248):\penalty0 820—823, June
	2009{\natexlab{a}}.
	\newblock ISSN 0028-0836.
	\newblock \doi{10.1038/nature08105}.
	
	\bibitem[Castro et~al.(2007)Castro, Novoselov, Morozov, Peres, dos Santos,
	Nilsson, Guinea, Geim, and Neto]{cite_5}
	Eduardo~V. Castro, K.~S. Novoselov, S.~V. Morozov, N.~M.~R. Peres, J.~M.
	B.~Lopes dos Santos, Johan Nilsson, F.~Guinea, A.~K. Geim, and A.~H.~Castro
	Neto.
	\newblock Biased bilayer graphene: Semiconductor with a gap tunable by the
	electric field effect.
	\newblock \emph{Phys. Rev. Lett.}, 99:\penalty0 216802, Nov 2007.
	\newblock \doi{10.1103/PhysRevLett.99.216802}.
	
	\bibitem[Okano et~al.(2020)Okano, Sharma, Ortmann, Nishimura, Günther, Gordan,
	Ikushima, Dzhagan, Salvan, and Zahn]{cite_6}
	Shun Okano, Apoorva Sharma, Frank Ortmann, Akira Nishimura, Christoph Günther,
	Ovidiu~D. Gordan, Kenji Ikushima, Volodymyr Dzhagan, Georgeta Salvan, and
	Dietrich R.~T. Zahn.
	\newblock Voltage-controlled dielectric function of bilayer graphene.
	\newblock \emph{Advanced Optical Materials}, 8\penalty0 (20):\penalty0 2000861,
	2020.
	\newblock \doi{10.1002/adom.202000861}.
	
	\bibitem[Ho et~al.(2010)Ho, Wu, Chiu, Wang, and Lin]{cite_7}
	Y.~H. Ho, J.~Y. Wu, Y.~H. Chiu, J.~Wang, and M.~F. Lin.
	\newblock Electronic and optical properties of monolayer and bilayer graphene.
	\newblock \emph{Philosophical Transactions of the Royal Society A:
		Mathematical, Physical and Engineering Sciences}, 368\penalty0
	(1932):\penalty0 5445--5458, 2010.
	\newblock \doi{10.1098/rsta.2010.0209}.
	
	\bibitem[Wang and Jin(2012)]{cite_8}
	Dali Wang and Guojun Jin.
	\newblock Tunable electronic transport characteristics through an aa-stacked
	bilayer graphene with magnetoelectric barriers.
	\newblock \emph{Journal of Applied Physics}, 112\penalty0 (5):\penalty0 053714,
	2012.
	\newblock \doi{10.1063/1.4751331}.
	
	\bibitem[Tran et~al.(2015)Tran, Lin, Glukhova, and Lin]{cite_9}
	Ngoc Thanh~Thuy Tran, Shih-Yang Lin, Olga~E. Glukhova, and Ming-Fa Lin.
	\newblock Configuration-induced rich electronic properties of bilayer graphene.
	\newblock \emph{The Journal of Physical Chemistry C}, 119\penalty0
	(19):\penalty0 10623--10630, 2015.
	\newblock \doi{10.1021/jp511692e}.
	
	\bibitem[Xu et~al.(2010)Xu, Li, and Dong]{cite_10}
	Yuehua Xu, Xiaowei Li, and Jinming Dong.
	\newblock Infrared and raman spectra of {AA}-stacking bilayer graphene.
	\newblock \emph{Nanotechnology}, 21\penalty0 (6):\penalty0 065711, jan 2010.
	\newblock \doi{10.1088/0957-4484/21/6/065711}.
	
	\bibitem[Abdullah et~al.(2018)Abdullah, Al~Ezzi, and Bahlouli]{cite_11}
	Hasan~M. Abdullah, Mohammed Al~Ezzi, and H.~Bahlouli.
	\newblock Electronic transport and klein tunneling in gapped aa-stacked bilayer
	graphene.
	\newblock \emph{Journal of Applied Physics}, 124\penalty0 (20):\penalty0
	204303, 2018.
	\newblock \doi{10.1063/1.5052402}.
	
	\bibitem[cit(2017)]{cite_12}
	\emph{2D Materials: Properties and Devices}.
	\newblock Cambridge University Press, 2017.
	\newblock \doi{10.1017/9781316681619}.
	
	\bibitem[Zhang et~al.(2009{\natexlab{b}})Zhang, Tang, Girit, Hao, Martin,
	Zettl, Crommie, Shen, and Wang]{cite_13}
	Yuanbo Zhang, Tsung-Ta Tang, Caglar Girit, Zhao Hao, Michael~C Martin, Alex
	Zettl, Michael~F Crommie, Y~Ron Shen, and Feng Wang.
	\newblock Direct observation of a widely tunable bandgap in bilayer graphene.
	\newblock \emph{Nature}, 459\penalty0 (7248), 2009{\natexlab{b}}.
	\newblock \doi{10.1038/nature08105}.
	
	\bibitem[Kotov et~al.(2012)Kotov, Uchoa, Pereira, Guinea, and
	Castro~Neto]{cite_14}
	Valeri~N. Kotov, Bruno Uchoa, Vitor~M. Pereira, F.~Guinea, and A.~H.
	Castro~Neto.
	\newblock Electron-electron interactions in graphene: Current status and
	perspectives.
	\newblock \emph{Rev. Mod. Phys.}, 84:\penalty0 1067--1125, Jul 2012.
	\newblock \doi{10.1103/RevModPhys.84.1067}.
	
	\bibitem[Liu et~al.(2009)Liu, Suenaga, Harris, and Iijima]{cite_15}
	Zheng Liu, Kazu Suenaga, Peter J.~F. Harris, and Sumio Iijima.
	\newblock Open and closed edges of graphene layers.
	\newblock \emph{Phys. Rev. Lett.}, 102:\penalty0 015501, Jan 2009.
	\newblock \doi{10.1103/PhysRevLett.102.015501}.
	
	\bibitem[Liu et~al.(2011)Liu, Ao, Wang, Wang, Sheng, and Yu]{cite_16}
	Yan Liu, Zhi-Min Ao, Tao Wang, Wen-Bo Wang, Kuang Sheng, and Bin Yu.
	\newblock Transformation from {AA} to {AB}-stacked bilayer graphene on
	$\alpha$-${SiO_2}$ under an electric field.
	\newblock \emph{Chinese Physics Letters}, 28\penalty0 (8):\penalty0 087303, aug
	2011.
	\newblock \doi{10.1088/0256-307x/28/8/087303}.
	
	\bibitem[Avsar et~al.(2020)Avsar, Ochoa, Guinea, \"Ozyilmaz, van Wees, and
	Vera-Marun]{cite_17}
	A.~Avsar, H.~Ochoa, F.~Guinea, B.~\"Ozyilmaz, B.~J. van Wees, and I.~J.
	Vera-Marun.
	\newblock Colloquium: Spintronics in graphene and other two-dimensional
	materials.
	\newblock \emph{Rev. Mod. Phys.}, 92:\penalty0 021003, Jun 2020.
	\newblock \doi{10.1103/RevModPhys.92.021003}.
	
	\bibitem[Laref et~al.(2020)Laref, Alsagri, e~Abbas, Laref, Huang, Xiong, Yang,
	Khandy, Rai, Varshney, and Wu]{cite_18}
	A.~Laref, M.~Alsagri, Syed Muhammad~Alay e~Abbas, S.~Laref, H.M. Huang, Y.C.
	Xiong, J.T. Yang, Shakeel~Ahmad Khandy, Dibya~Prakash Rai, Dinesh Varshney,
	and Xiaozhi Wu.
	\newblock Electronic structure and optical characteristics of aa stacked
	bilayer graphene: A first principles calculations.
	\newblock \emph{Optik}, 206:\penalty0 163755, 2020.
	\newblock ISSN 0030-4026.
	\newblock \doi{10.1016/j.ijleo.2019.163755}.
	
	\bibitem[Ju et~al.(2017)Ju, Wang, Cao, Taniguchi, Watanabe, Louie, Rana, Park,
	Hone, Wang, and McEuen]{cite_19}
	Long Ju, Lei Wang, Ting Cao, Takashi Taniguchi, Kenji Watanabe, Steven~G.
	Louie, Farhan Rana, Jiwoong Park, James Hone, Feng Wang, and Paul~L. McEuen.
	\newblock Tunable excitons in bilayer graphene.
	\newblock \emph{Science}, 358\penalty0 (6365):\penalty0 907--910, 2017.
	\newblock ISSN 0036-8075.
	\newblock \doi{10.1126/science.aam9175}.
	
	\bibitem[{Li} et~al.(2017){Li}, {Taniguchi}, {Watanabe}, {Hone}, and
	{Dean}]{cite_20}
	J.~I.~A. {Li}, T.~{Taniguchi}, K.~{Watanabe}, J.~{Hone}, and C.~R. {Dean}.
	\newblock {Excitonic superfluid phase in double bilayer graphene}.
	\newblock \emph{Nature Physics}, 13\penalty0 (8):\penalty0 751--755, August
	2017.
	\newblock \doi{10.1038/nphys4140}.
	
	\bibitem[Neilson and Peeters(2014)]{cite_21}
	D.~Neilson and F.~Peeters.
	\newblock Evidence of high-temperature exciton condensation in two-dimensional
	atomic double layers.
	\newblock \emph{Scientific Reports}, 4\penalty0 (1), 2014.
	\newblock \doi{10.1038/srep07319}.
	
	\bibitem[Wang et~al.(2020)Wang, Nie, Li, Zuo, Fauqu{\'e}, Zhu, and
	Behnia]{cite_22}
	Jinhua Wang, Pan Nie, Xiaokang Li, Huakun Zuo, Beno{\^\i}t Fauqu{\'e}, Zengwei
	Zhu, and Kamran Behnia.
	\newblock Critical point for bose{\textendash}einstein condensation of excitons
	in graphite.
	\newblock \emph{Proceedings of the National Academy of Sciences}, 117\penalty0
	(48):\penalty0 30215--30219, 2020.
	\newblock ISSN 0027-8424.
	\newblock \doi{10.1073/pnas.2012811117}.
	
	\bibitem[Kharitonov and Efetov(2008)]{cite_23}
	Maxim~Yu. Kharitonov and Konstantin~B. Efetov.
	\newblock Electron screening and excitonic condensation in double-layer
	graphene systems.
	\newblock \emph{Phys. Rev. B}, 78:\penalty0 241401, Dec 2008.
	\newblock \doi{10.1103/PhysRevB.78.241401}.
	
	\bibitem[Min et~al.(2008)Min, Bistritzer, Su, and MacDonald]{cite_24}
	Hongki Min, Rafi Bistritzer, Jung-Jung Su, and A.~H. MacDonald.
	\newblock Room-temperature superfluidity in graphene bilayers.
	\newblock \emph{Phys. Rev. B}, 78:\penalty0 121401, Sep 2008.
	\newblock \doi{10.1103/PhysRevB.78.121401}.
	
	\bibitem[Fogler et~al.(2014)Fogler, Butov, and Novoselov]{cite_25}
	MM~Fogler, LV~Butov, and KS~Novoselov.
	\newblock High-temperature superfluidity with indirect excitons in van der
	waals heterostructures.
	\newblock \emph{Nature communications}, 5:\penalty0 4555, 2014.
	\newblock ISSN 2041-1723.
	\newblock \doi{10.1038/ncomms5555}.
	
	\bibitem[Apinyan and Kopeć(2018)]{cite_26}
	V.~Apinyan and T.K. Kopeć.
	\newblock Spectral properties of excitons in the bilayer graphene.
	\newblock \emph{Physica E: Low-dimensional Systems and Nanostructures},
	95:\penalty0 108--120, 2018.
	\newblock ISSN 1386-9477.
	\newblock \doi{10.1016/j.physe.2017.09.015}.
	
	\bibitem[Akzyanov et~al.(2014)Akzyanov, Sboychakov, Rozhkov, Rakhmanov, and
	Nori]{cite_27}
	R.~S. Akzyanov, A.~O. Sboychakov, A.~V. Rozhkov, A.~L. Rakhmanov, and Franco
	Nori.
	\newblock $aa$-stacked bilayer graphene in an applied electric field: Tunable
	antiferromagnetism and coexisting exciton order parameter.
	\newblock \emph{Phys. Rev. B}, 90:\penalty0 155415, Oct 2014.
	\newblock \doi{10.1103/PhysRevB.90.155415}.
	
	\bibitem[Rakhmanov et~al.(2012)Rakhmanov, Rozhkov, Sboychakov, and
	Nori]{cite_28}
	A.~L. Rakhmanov, A.~V. Rozhkov, A.~O. Sboychakov, and Franco Nori.
	\newblock Instabilities of the $aa$-stacked graphene bilayer.
	\newblock \emph{Phys. Rev. Lett.}, 109:\penalty0 206801, Nov 2012.
	\newblock \doi{10.1103/PhysRevLett.109.206801}.
	
	\bibitem[Sboychakov et~al.(2021)Sboychakov, Rakhmanov, Rozhkov, and
	Nori]{cite_29}
	A.~O. Sboychakov, A.~L. Rakhmanov, A.~V. Rozhkov, and Franco Nori.
	\newblock Bilayer graphene can become a fractional metal.
	\newblock \emph{Phys. Rev. B}, 103:\penalty0 L081106, Feb 2021.
	\newblock \doi{10.1103/PhysRevB.103.L081106}.
	
	\bibitem[Szałowski(2017)]{cite_30}
	Karol Szałowski.
	\newblock Ferrimagnetic and antiferromagnetic phase in bilayer graphene
	nanoflake controlled with external electric fields.
	\newblock \emph{Carbon}, 118:\penalty0 78--85, 2017.
	\newblock ISSN 0008-6223.
	\newblock \doi{10.1016/j.carbon.2017.03.019}.
	
	\bibitem[Wang et~al.(2012)Wang, Guo, Liu, and Sheng]{cite_31}
	Tao Wang, Qing Guo, Yan Liu, and Kuang Sheng.
	\newblock A comparative investigation of an {AB}- and {AA}-stacked bilayer
	graphene sheet under an applied electric field: A density functional theory
	study.
	\newblock \emph{Chinese Physics B}, 21\penalty0 (6):\penalty0 067301, jun 2012.
	\newblock \doi{10.1088/1674-1056/21/6/067301}.
	
	\bibitem[Jadaun et~al.(2013)Jadaun, Movva, Register, and Banerjee]{cite_32}
	Priyamvada Jadaun, Hema C.~P. Movva, Leonard~F. Register, and Sanjay~K.
	Banerjee.
	\newblock Theory and synthesis of bilayer graphene intercalated with icl and
	ibr for low power device applications.
	\newblock \emph{Journal of Applied Physics}, 114\penalty0 (6):\penalty0 063702,
	2013.
	\newblock \doi{10.1063/1.4817498}.
	
	\bibitem[Roy et~al.(1998)Roy, Kallinger, and Sattler]{cite_33}
	H.-V. Roy, C.~Kallinger, and K.~Sattler.
	\newblock Study of single and multiple foldings of graphitic sheets.
	\newblock \emph{Surface Science}, 407\penalty0 (1):\penalty0 1--6, 1998.
	\newblock ISSN 0039-6028.
	\newblock \doi{10.1016/S0039-6028(97)01032-7}.
	
	\bibitem[Borysiuk et~al.(2011)Borysiuk, Sołtys, and Piechota]{cite_34}
	J.~Borysiuk, J.~Sołtys, and J.~Piechota.
	\newblock Stacking sequence dependence of graphene layers on sic:experimental
	and theoretical investigation.
	\newblock \emph{Journal of Applied Physics}, 109\penalty0 (9):\penalty0 093523,
	2011.
	\newblock \doi{10.1063/1.3585829}.
	
	\bibitem[Lee et~al.(2008)Lee, Lee, Ahn, Kim, Wilson, and John]{cite_35}
	Jae-Kap Lee, Seung-Cheol Lee, Jae-Pyoung Ahn, Soo-Chul Kim, John I.~B. Wilson,
	and Phillip John.
	\newblock The growth of aa graphite on (111) diamond.
	\newblock \emph{The Journal of Chemical Physics}, 129\penalty0 (23):\penalty0
	234709, 2008.
	\newblock \doi{10.1063/1.2975333}.
	
	\bibitem[Liu et~al.(2014)Liu, Liu, Wang, and Ao]{cite_36}
	Hai-Long Liu, Yan Liu, Tao Wang, and Zhi-Min Ao.
	\newblock {AA} bilayer graphene on si-terminated {SiO}2 under electric field.
	\newblock \emph{Chinese Physics B}, 23\penalty0 (2):\penalty0 026802, feb 2014.
	\newblock \doi{10.1088/1674-1056/23/2/026802}.
	
	\bibitem[de~Freitas et~al.(2017)de~Freitas, Sanz, and Villas-B\^oas]{cite_37}
	Antonio de~Freitas, L.~Sanz, and Jos\'e~M. Villas-B\^oas.
	\newblock Coherent control of the dynamics of a single quantum-dot exciton
	qubit in a cavity.
	\newblock \emph{Phys. Rev. B}, 95:\penalty0 115110, Mar 2017.
	\newblock \doi{10.1103/PhysRevB.95.115110}.
	
	\bibitem[Michaelis~de Vasconcellos et~al.(2010)Michaelis~de Vasconcellos,
	Gordon, Bichler, Meier, and Zrenner]{cite_38}
	S.~Michaelis~de Vasconcellos, S.~Gordon, M.~Bichler, T.~Meier, and A.~Zrenner.
	\newblock Coherent control of a single exciton qubit by optoelectronic
	manipulation.
	\newblock \emph{Nature Photonics}, 4\penalty0 (8):\penalty0 545–548, 2010.
	\newblock \doi{10.1038/nphoton.2010.124}.
	
	\bibitem[Apinyan and Kopeć(2020)]{cite_39}
	V.~Apinyan and T.K. Kopeć.
	\newblock High thermoelectric performance in excitonic bilayer graphene.
	\newblock \emph{Physica E: Low-dimensional Systems and Nanostructures},
	124:\penalty0 114234, 2020.
	\newblock ISSN 1386-9477.
	\newblock \doi{10.1016/j.physe.2020.114234}.
	
	\bibitem[Gordon et~al.(2010)Gordon, Bichler, Meier, and Zrenner]{cite_40}
	S.~Gordon, M.~Bichler, T.~Meier, and A.~Zrenner.
	\newblock Sensitization of silicon by singlet exciton fission in tetracene.
	\newblock \emph{Nature Photonics}, 4\penalty0 (8):\penalty0 545–548, 2010.
	\newblock \doi{10.1038/nphoton.2010.124}.
	
	\bibitem[Jiang et~al.(2020)Jiang, Lou, Liu, Li, Song, Chang, Duan, and
	Zhang]{cite_41}
	Zeyu Jiang, Wenkai Lou, Yu~Liu, Yuanchang Li, Haifeng Song, Kai Chang, Wenhui
	Duan, and Shengbai Zhang.
	\newblock Spin-triplet excitonic insulator: The case of semihydrogenated
	graphene.
	\newblock \emph{Phys. Rev. Lett.}, 124:\penalty0 166401, Apr 2020.
	\newblock \doi{10.1103/PhysRevLett.124.166401}.
	
	\bibitem[Zelezny et~al.(2018)Zelezny, Wadley, Olejnik, Hoffmann, and
	Ohno]{cite_42}
	J.~Zelezny, P.~Wadley, K.~Olejnik, A.~Hoffmann, and H.~Ohno.
	\newblock Spin transport and spin torque in antiferromagnetic devices.
	\newblock \emph{Nature Physics}, 14\penalty0 (8):\penalty0 220–228, 2018.
	\newblock \doi{10.1038/s41567-018-0062-7}.
	
	\bibitem[Yan et~al.(2020)Yan, Feng, Qin, Zhou, Guo, Wang, Chen, Zhang, Wu,
	Jiang, and Liu]{cite_43}
	Han Yan, Zexin Feng, Peixin Qin, Xiaorong Zhou, Huixin Guo, Xiaoning Wang,
	Hongyu Chen, Xin Zhang, Haojiang Wu, Chengbao Jiang, and Zhiqi Liu.
	\newblock Electric-field-controlled antiferromagnetic spintronic devices.
	\newblock \emph{Advanced Materials}, 32\penalty0 (12):\penalty0 1905603, 2020.
	\newblock \doi{10.1002/adma.201905603}.
	
	\bibitem[Sboychakov et~al.(2013{\natexlab{a}})Sboychakov, Rozhkov, Rakhmanov,
	and Nori]{cite_44}
	A.~O. Sboychakov, A.~V. Rozhkov, A.~L. Rakhmanov, and Franco Nori.
	\newblock Antiferromagnetic states and phase separation in doped aa-stacked
	graphene bilayers.
	\newblock \emph{Phys. Rev. B}, 88:\penalty0 045409, Jul 2013{\natexlab{a}}.
	\newblock \doi{10.1103/PhysRevB.88.045409}.
	
	\bibitem[Sboychakov et~al.(2013{\natexlab{b}})Sboychakov, Rakhmanov, Rozhkov,
	and Nori]{cite_45}
	A.~O. Sboychakov, A.~L. Rakhmanov, A.~V. Rozhkov, and Franco Nori.
	\newblock Metal-insulator transition and phase separation in doped aa-stacked
	graphene bilayer.
	\newblock \emph{Phys. Rev. B}, 87:\penalty0 121401, Mar 2013{\natexlab{b}}.
	\newblock \doi{10.1103/PhysRevB.87.121401}.
	
	\bibitem[Wehling et~al.(2011)Wehling, \ifmmode \mbox{\c{S}}\else
	\c{S}\fi{}a\ifmmode \mbox{\c{s}}\else \c{s}\fi{}\ifmmode \imath \else \i
	\fi{}o\ifmmode~\breve{g}\else \u{g}\fi{}lu, Friedrich, Lichtenstein,
	Katsnelson, and Bl\"ugel]{cite_46}
	T.~O. Wehling, E.~\ifmmode \mbox{\c{S}}\else \c{S}\fi{}a\ifmmode
	\mbox{\c{s}}\else \c{s}\fi{}\ifmmode \imath \else \i
	\fi{}o\ifmmode~\breve{g}\else \u{g}\fi{}lu, C.~Friedrich, A.~I. Lichtenstein,
	M.~I. Katsnelson, and S.~Bl\"ugel.
	\newblock Strength of effective coulomb interactions in graphene and graphite.
	\newblock \emph{Phys. Rev. Lett.}, 106:\penalty0 236805, Jun 2011.
	\newblock \doi{10.1103/PhysRevLett.106.236805}.
	
	\bibitem[Sorella and Tosatti(1992)]{cite_47}
	S.~Sorella and E.~Tosatti.
	\newblock Semi-metal-insulator transition of the hubbard model in the honeycomb
	lattice.
	\newblock \emph{Europhysics Letters ({EPL})}, 19\penalty0 (8):\penalty0
	699--704, aug 1992.
	\newblock \doi{10.1209/0295-5075/19/8/007}.
	
	\bibitem[Martelo et~al.(1996)Martelo, Dzierzawa, Siffert, and
	Baeriswyl]{cite_48}
	L.M. Martelo, M.~Dzierzawa, L.~Siffert, and D.~Baeriswyl.
	\newblock Mott-hubbard transition and antiferromagnetism on the honeycomb
	lattice.
	\newblock \emph{Zeitschrift für Physik B Condensed Matter}, 103\penalty0
	(2):\penalty0 335–338, 1996.
	\newblock \doi{10.1038/nphoton.2010.124}.
	
	\bibitem[Paiva et~al.(2005)Paiva, Scalettar, Zheng, Singh, and Oitmaa]{cite_49}
	Thereza Paiva, R.~T. Scalettar, W.~Zheng, R.~R.~P. Singh, and J.~Oitmaa.
	\newblock Ground-state and finite-temperature signatures of quantum phase
	transitions in the half-filled hubbard model on a honeycomb lattice.
	\newblock \emph{Phys. Rev. B}, 72:\penalty0 085123, Aug 2005.
	\newblock \doi{10.1103/PhysRevB.72.085123}.
	
	\bibitem[Nikolaev and Ulybyshev(2014)]{cite_50}
	A.~Nikolaev and M.~Ulybyshev.
	\newblock Monte-carlo study of the phase transition in the aa-stacked bilayer
	graphene.
	\newblock \emph{Proceedings of Science}, 214\penalty0 (8), 2014.
	\newblock \doi{10.22323/1.214.0054}.
	
	\bibitem[Blom(2005)]{cite_51}
	A~Blom.
	\newblock Exact solution of the zeeman effect in single-electron systems.
	\newblock \emph{Physica Scripta}, T120:\penalty0 90--98, jan 2005.
	\newblock \doi{10.1088/0031-8949/2005/t120/014}.
	
	\bibitem[Aoki and S.~Dresselhaus(2014)]{cite_52}
	Hideo Aoki and Mildred S.~Dresselhaus, editors.
	\newblock Springer International Publishing, Cham, 2014.
	\newblock ISBN 978-3-319-02632-9.
	\newblock \doi{10.1007/978-3-319-02633-6}.
	
	\bibitem[Negele and Orland(1998)]{cite_53}
	John~W. Negele and Henri Orland.
	\newblock \emph{Quantum Many-particle Systems}.
	\newblock Westview Press, 1998.
	\newblock ISBN 0738200522.
	
	\bibitem[Apinyan and Kope\'{c}(2008)]{cite_54}
	V.A. Apinyan and T.K. Kope\'{c}.
	\newblock Effective pairing interaction in the two-dimensional hubbard model
	within a spin rotationally invariant approach.
	\newblock \emph{Phys. Rev. B}, 78:\penalty0 184511, Nov 2008.
	\newblock \doi{10.1103/PhysRevB.78.184511}.
	
	\bibitem[A.A. et~al.(1963)A.A., L.P., and I.E.]{cite_55}
	Abrikosov A.A., Gorkov L.P., and Dzyaloshinski I.E.
	\newblock \emph{Methods of quantum field theory in statistical physics}.
	\newblock Dover, New York, N.Y., 1963.
	
	\bibitem[cit(2010)]{cite_56}
	Room-temperature antiferromagnetic memory resistor.
	\newblock \emph{Nature Photonics}, 4\penalty0 (8):\penalty0 545–548, 2010.
	\newblock \doi{10.1038/nphoton.2010.124}.
	
	\bibitem[Wang et~al.(2017)Wang, Song, Zhang, and Pan]{cite_57}
	Y.Y. Wang, C.~Song, J.Y. Zhang, and F.~Pan.
	\newblock Spintronic materials and devices based on antiferromagnetic metals.
	\newblock \emph{Progress in Natural Science: Materials International},
	27\penalty0 (2):\penalty0 208--216, 2017.
	\newblock ISSN 1002-0071.
	\newblock \doi{10.1016/j.pnsc.2017.03.008}.
	
	\bibitem[Jenkins et~al.(2019)Jenkins, Meo, Elliott, Piotrowski, Bapna,
	Chantrell, Majetich, and Evans]{cite_58}
	Sarah Jenkins, Andrea Meo, Luke~E Elliott, Stephan~K Piotrowski, Mukund Bapna,
	Roy~W Chantrell, Sara~A Majetich, and Richard F~L Evans.
	\newblock Magnetic stray fields in nanoscale magnetic tunnel junctions.
	\newblock \emph{Journal of Physics D: Applied Physics}, 53\penalty0
	(4):\penalty0 044001, nov 2019.
	\newblock \doi{10.1088/1361-6463/ab4fbf}.
	
\end{thebibliography}
\end{document}